\documentclass[pdflatex,sn-nature]{sn-jnl}

\usepackage{graphicx}
\graphicspath{{figures/}}
\usepackage{placeins}
\usepackage{multirow}
\usepackage{amsmath,amssymb,amsfonts}
\usepackage{booktabs}
\usepackage{xcolor}
\usepackage{algorithm}
\usepackage{algorithmicx}
\usepackage{algpseudocode}
\usepackage{xurl}

\makeatletter
\@ifundefined{@makespecialcolbox}{%
  \@ifundefined{@make@specialcolbox}{}{%
    \let\@makespecialcolbox\@make@specialcolbox
  }%
}{}
\def\toclevel@paragraph{3}
\def\toclevel@subparagraph{1}
\makeatother

\begin{document}

\title[Self-improving Data Synthesis]{Self-improving data synthesis can scale embedding-model adaptation across domains and tasks}

\author[1,2]{\fnm{Zucan} \sur{Lyu}}\email{zucanlv@mail.ustc.edu.cn}
\equalcont{These authors contributed equally: Zucan Lyu, Rui Tang, Jianlyu Chen.}
\author[1,2]{\fnm{Rui} \sur{Tang}}
\equalcont{These authors contributed equally: Zucan Lyu, Rui Tang, Jianlyu Chen.}
\author[1,2]{\fnm{Jianlyu} \sur{Chen}}
\equalcont{These authors contributed equally: Zucan Lyu, Rui Tang, Jianlyu Chen.}
\author[1,2]{\fnm{Peisheng} \sur{Jin}}
\author*[1,2]{\fnm{Defu} \sur{Lian}}\email{liandefu@ustc.edu.cn}

\affil[1]{\orgname{University of Science and Technology of China}, \orgaddress{\city{Hefei}, \country{China}}}
\affil[2]{\orgname{State Key Laboratory of Cognitive Intelligence}, \orgaddress{\city{Hefei}, \country{China}}}

% NMI Articles use an unreferenced abstract of at most 150 words.
\abstract{

AI agents require specialised retrieval across domains and tasks, but adapting embedding models is constrained by costly manual training-data design. Here we introduce SYNTRA, a self-improving agentic data synthesis framework that turns a few retrieval examples into reusable synthesis instructions. Given an unlabelled corpus and a few relevant query-document pairs, SYNTRA infers implicit information needs and relevance criteria, then uses feedback on trial outputs to refine instructions for query generation and relevance assessment. Experiments across 24 datasets spanning 15 domains demonstrate that the same synthesis workflow produces effective training data, benefiting even strong embedding models with and without existing labelled training sets. Self-improvement strengthens this supervision on all six evaluated tasks, yielding better retrieval than human calibration on four. The training value persists at larger scales: on MS MARCO, retrieval gains continue across a 32-fold range of synthetic training-data sizes. SYNTRA thus enables scalable embedding-model adaptation with minimal human intervention.

}


\maketitle
% ============================================================================
% Introduction
% NMI requires the Introduction to appear without an ``Introduction'' heading.
% ============================================================================

\begin{figure}[!htbp]
\centering
\includegraphics[width=\textwidth,height=0.66\textheight,keepaspectratio]{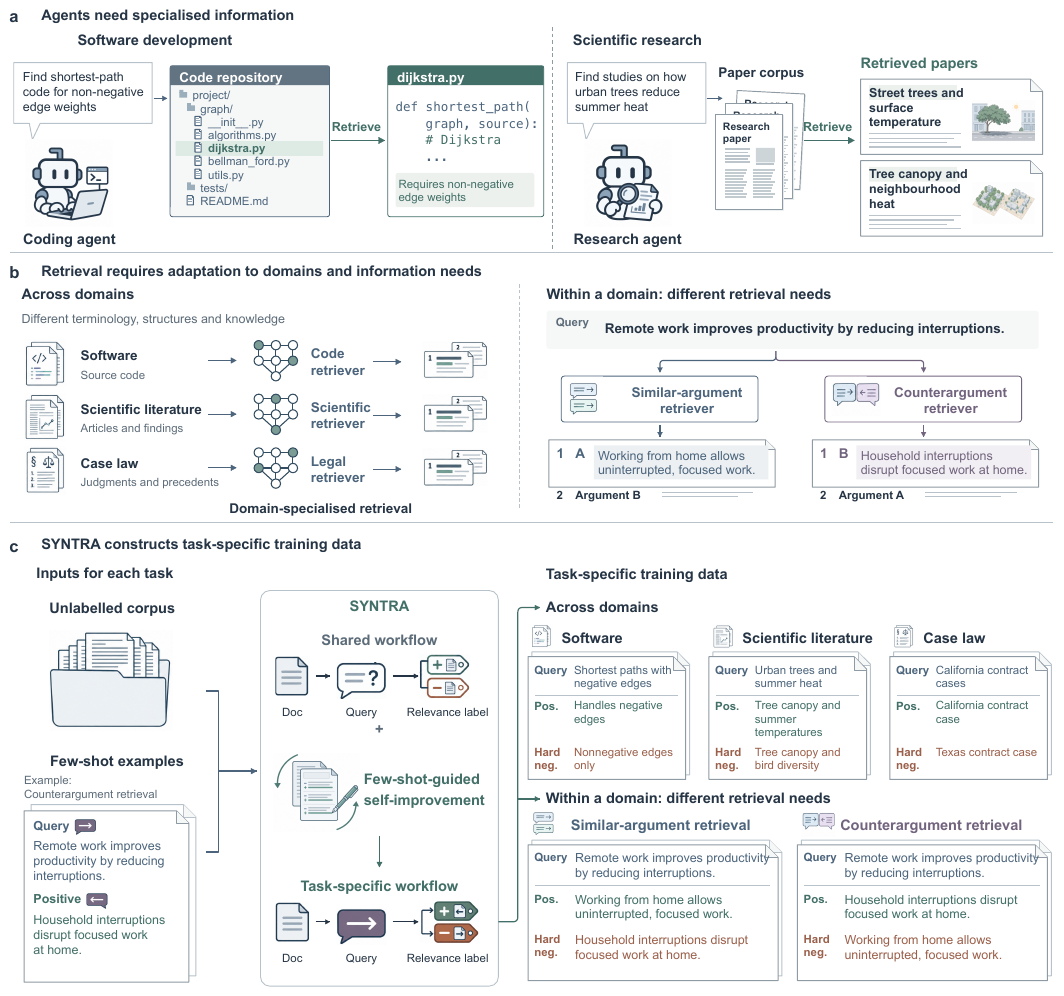}
\caption{\textbf{Constructing task-specific training data for specialised retrieval.}
\textbf{a}, A coding agent searches a repository and retrieves the implementation that satisfies a stated constraint. A research agent retrieves supporting studies from a paper corpus.
\textbf{b}, Documents differ across domains in terminology, structure and knowledge. Within a single domain, the user's objective changes which documents are relevant. For the same query, similar-argument and counterargument retrieval rank the same two arguments in opposite orders.
\textbf{c}, SYNTRA receives an unlabelled corpus and a small set of few-shot examples, each pairing a query with a positive document. Few-shot-guided self-improvement turns a shared synthesis workflow into a task-specific one. This workflow then constructs (query, positive, negative) training records at scale. SYNTRA is run separately for each retrieval task using the same framework. The examples shown span software, scientific literature and case law. Within a domain, the same two arguments exchange positive and hard-negative roles under different retrieval needs. Pos., positive document; Hard neg., hard-negative document. All examples are schematic and abbreviated, not generated samples or measured retrieval results.}
\label{fig:syntra-concept}
\end{figure}

Retrieval is becoming a core capability of artificial intelligence (AI) agents, providing access to external knowledge that informs reasoning and action. For example, Cursor uses a specialised embedding model alongside conventional search tools to locate relevant code (\url{https://cursor.com/blog/semsearch}), while OpenScholar retrieves scientific evidence to answer research questions \cite{asai2026openscholar}. These applications require retrieval tailored to domain knowledge and task-specific information needs (Fig.~\ref{fig:syntra-concept}a). As agents are deployed across more domains and tasks, the demand for specialised retrieval grows.

Meeting this demand requires adaptation to both new corpora and task-specific relevance criteria. New corpora require retrieval systems to adapt to unfamiliar content. During the COVID-19 pandemic, for example, the fast-growing scientific literature created search needs that existing collections did not cover, and TREC-COVID was organised to evaluate how well retrieval systems could serve them \cite{roberts2020treccovid}. Yet adaptation concerns not only which documents are available but also which ones count as relevant: different user objectives can make the same document relevant to one task and unsuitable for another, even within the same corpus \cite{song2025ifir} (Fig.~\ref{fig:syntra-concept}b). Embedding models must therefore capture both domain knowledge and what constitutes relevant information for each task. The challenge is to make this adaptation scalable as domains and retrieval needs multiply.

A key bottleneck in this adaptation is constructing training data that reflect domain knowledge and task-specific relevance criteria. Unlabelled corpora are often available, but task-specific supervision remains scarce \cite{wang2022gpl,li2024meticulous}. To learn task-relevant distinctions, embedding models need training examples that associate queries with relevant documents and informative negatives \cite{karpukhin2020dpr,zhou2022simans}. Constructing these examples requires deciding what users might ask, which documents satisfy their needs and which plausible alternatives fail to do so. These decisions must be revisited as domains and tasks change, making training-data construction a recurring obstacle to scalable retrieval adaptation.

\enlargethispage{2\baselineskip}
Large language models (LLMs) offer a way to turn unlabelled corpora into retrieval training data \cite{ma2020zeroshot,jeronymo2023inparsv2}. Their ability to generate queries \cite{bonifacio2022inpars,dai2023promptagator,xian2024path,saadfalcon2023udapdr,mayfield2023synthetic,abonizio2023inparstoolkit,fang2024scalingdense,wen2025synthetic,chaudhary2024relative} and assess document relevance \cite{lee2024gecko,thomas2024searcherpreferences,sun2023chatgpt} has enabled effective training datasets for general-purpose retrieval \cite{wang2024improving,lee2024gecko,almeida2024exploring,chen2025littlegiants,krastev2025inparsplus,springer2025influence}, code retrieval \cite{li2025coder} and reasoning-intensive document retrieval \cite{shao2025reasonir,chen2026reasonembed}. However, applying these capabilities to a target task can still depend on task-specific human input, including defining information needs and relevance criteria, screening proposed tasks, and refining generation and annotation instructions \cite{dai2023promptagator,mayfield2023synthetic,wen2025synthetic,li2025coder,chen2026reasonembed}. Repeating this design work across domains and tasks makes human effort a bottleneck to scalable retrieval adaptation. This raises a central question: can a data synthesis framework infer task requirements and refine its own instructions with minimal human intervention, enabling embedding-model adaptation to scale across domains and tasks?

Here we introduce SYNTRA, a self-improving agentic data synthesis framework that turns a few retrieval examples into reusable data synthesis instructions. Given an unlabelled corpus and a small set of relevant query-document pairs, SYNTRA infers the task-specific information needs and relevance criteria implicit in the examples and develops initial synthesis instructions. Through small-sample generate-evaluate-refine cycles, it uses feedback on trial outputs to refine instructions governing query generation and relevance assessment. These revisions make task requirements explicit, guiding the construction of queries, identification of relevant documents and distinction of hard negatives. Once refined, these task-specific instructions are shared across documents and remain fixed throughout large-scale data production (Fig.~\ref{fig:syntra-concept}c). By converting few-shot guidance into reusable instructions, SYNTRA reduces the need for manual synthesis refinement, enabling a shared workflow to support scalable embedding-model adaptation across domains and tasks.

Experiments across 24 retrieval datasets spanning 15 domains demonstrate that the same SYNTRA workflow produces effective task-specific training data across diverse retrieval settings. These data also benefit strong embedding models, supplying supervision where labelled training sets are unavailable and complementing existing supervision where they are available. Self-improvement strengthens this supervision on all six evaluated tasks, yielding better downstream retrieval than human calibration on four. The training value also persists at larger scales: on MS MARCO, retrieval performance improves at every evaluated scale across a 32-fold range of synthetic training-data sizes. Together, these findings support scalable embedding-model adaptation with minimal human intervention, which could help meet the growing demand for specialised retrieval as AI agents are deployed across more domains and tasks.

\section*{The SYNTRA framework}

\begin{figure}[!htbp]
\centering
\includegraphics[width=\textwidth]{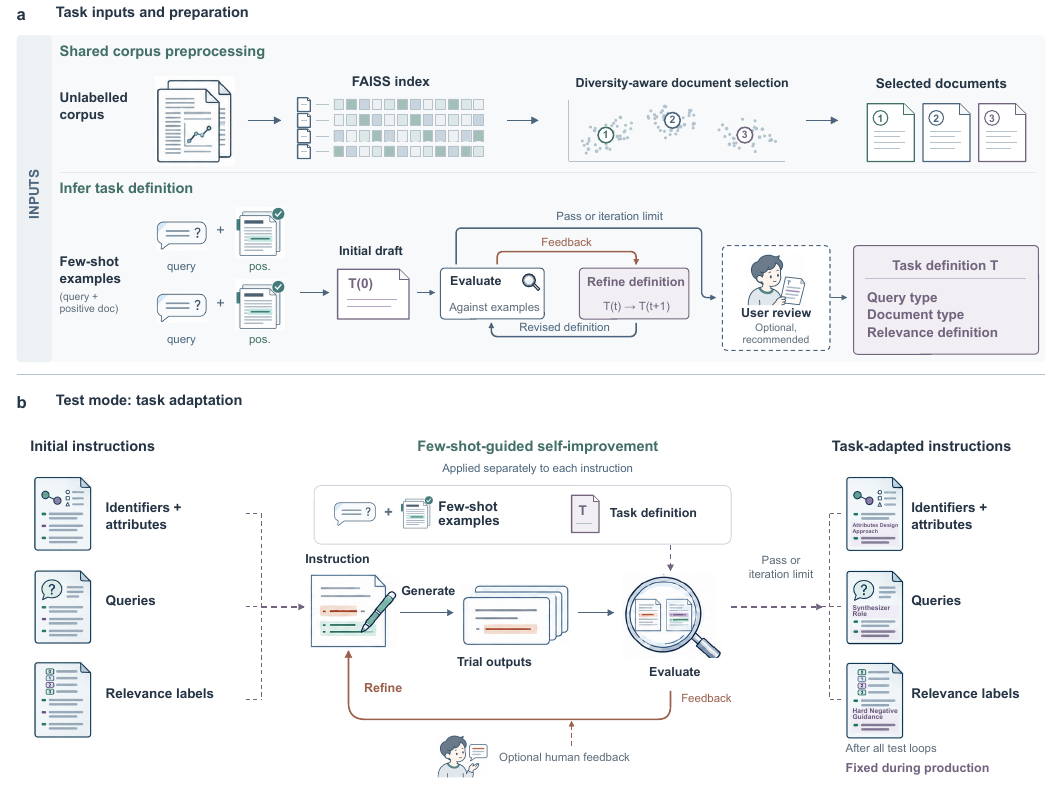}
\caption{\textbf{Task preparation and adaptation in SYNTRA.}
\textbf{a}, SYNTRA receives an unlabelled corpus and a small set of few-shot examples pairing queries with positive documents. Corpus preprocessing builds a full-corpus index and selects diverse source documents. The task definition is inferred from the few-shot examples; optional user review is recommended. Point groups schematically illustrate semantic diversity, not clustering outputs.
\textbf{b}, Small-sample generate-evaluate-refine cycles establish reusable task-adapted instructions before production. Each test loop ends when verification passes or its iteration limit is reached. The shared instructions then remain fixed throughout production.}
\label{fig:syntra-overview}
\end{figure}

\begin{figure}[!htbp]
\centering
\includegraphics[width=\textwidth]{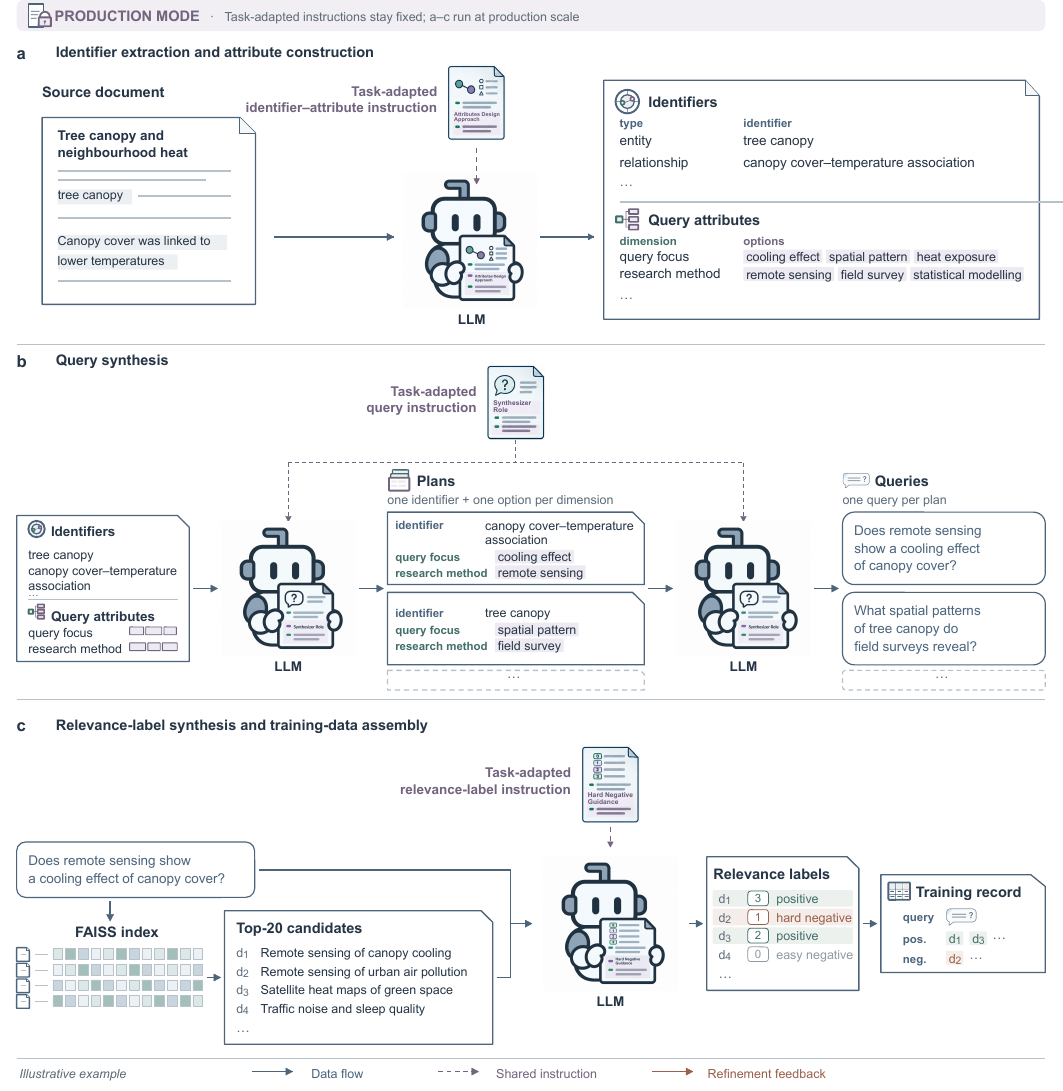}
\caption{\textbf{Production-scale synthesis in SYNTRA.}
Task-adapted instructions established in test mode remain fixed during production.
\textbf{a}, For each selected document, an LLM jointly extracts semantic identifiers and constructs query attributes with dimensions and options.
\textbf{b}, Query synthesis forms plans that pair one semantic identifier with one option from each query-attribute dimension. It then writes one query per plan using the same task-adapted instruction for both steps.
\textbf{c}, Relevance annotation retrieves candidates from the full corpus and labels them with the task-adapted relevance annotation instruction. The labelled candidates are assembled into (query, positive, negative) training records. Document content, plans, queries, candidates and labels are illustrative.}
\label{fig:syntra-production}
\end{figure}

As shown in Figs.~\ref{fig:syntra-overview} and \ref{fig:syntra-production}, SYNTRA adapts the same retrieval data synthesis workflow to different tasks through a self-improvement loop that refines task-specific synthesis instructions. In \emph{test mode}, it infers task requirements from a few relevant query-document examples and uses these requirements to guide iterative cycles of trial generation, evaluation and instruction refinement. In \emph{production mode}, the refined instructions remain fixed and are reused across documents within each task to construct training data at scale.

The user supplies an unlabelled corpus and a small set of examples pairing queries with relevant documents (Fig.~\ref{fig:syntra-overview}a). Corpus preprocessing selects diverse source documents for query generation and builds a full-corpus index for candidate retrieval. From the examples, SYNTRA infers a task definition specifying the intended query type, document type and relevance criteria. It evaluates and refines this definition against the examples, establishing common task requirements for the synthesis modules. The resulting definition guides the initial instructions for query synthesis and relevance annotation.

In \emph{test mode}, SYNTRA evaluates whether its trial outputs conform to the task definition and the patterns conveyed by the provided examples (Fig.~\ref{fig:syntra-overview}b). It uses the resulting diagnostic feedback to revise its synthesis instructions, refining how queries express information needs and how documents are assessed for relevance. SYNTRA then generates and evaluates new trial outputs using the revised instructions, forming a self-improvement loop that continues until verification passes or the iteration limit is reached. Once the loops finish, the refined instructions remain fixed and are reused across documents, allowing the feedback-driven refinements made during testing to guide production-scale synthesis.

In \emph{production mode}, SYNTRA uses the task-specific instructions refined through self-improvement to guide a three-stage workflow for large-scale training data synthesis. To generate queries that express diverse information needs grounded in the corpus, identifier extraction and attribute construction identify what users might seek within a document and how those needs can vary (Fig.~\ref{fig:syntra-production}a). Query synthesis uses these components to plan information needs and express them as queries (Fig.~\ref{fig:syntra-production}b). Relevance annotation retrieves candidate documents from the full corpus and assesses them against the task-specific relevance criteria, distinguishing positives from hard and easy negatives (Fig.~\ref{fig:syntra-production}c). The labelled query-document pairs are then assembled into training records for embedding-model adaptation.

% ============================================================================
\section{Results}\label{sec:results}
% ============================================================================

\subsection*{One SYNTRA workflow delivers consistent retrieval gains across domains and tasks}\label{subsec:rq3}

\begin{figure}[!htbp]
\centering
\includegraphics[width=\textwidth]{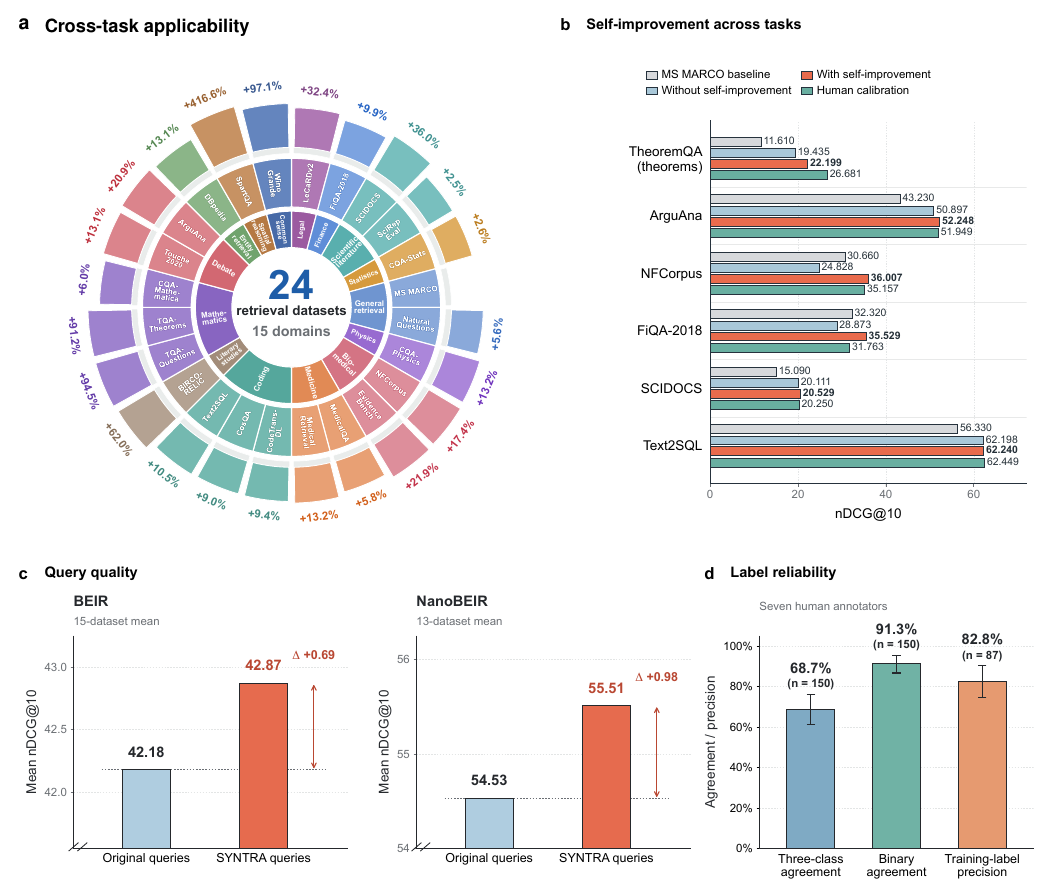}
\caption{\textbf{Cross-task applicability, self-improvement and quality of SYNTRA-generated supervision.}
\textbf{a}, Coverage of 24 datasets spanning 15 domains, with relative retrieval gains for the 23 task-specific comparisons against the MS MARCO-trained baseline. Numeric labels report percentage gains; radial extensions use a nonlinear scale. The CodeTrans-DL gain averages three reported variants. MS MARCO is included as the common training corpus without a task-adaptation gain.
\textbf{b}, Retrieval performance of the MS MARCO-trained baseline and models trained on data generated without self-improvement, with automatic self-improvement or with human calibration. The six tasks were TheoremQA (theorems), ArguAna, NFCorpus, FiQA-2018 (FiQA), SCIDOCS and Text2SQL.
\textbf{c}, Mean nDCG@10 on BEIR and NanoBEIR for models trained with original or SYNTRA-generated queries under matched conditions.
\textbf{d}, Agreement between Qwen3-30B-A3B relevance labels and reference labels derived from votes by seven human annotators. Binary and three-class agreement were evaluated on 150 query-document pairs, with 50 pairs from each of three tasks. Binary evaluation distinguishes relevant from non-relevant documents; three-class evaluation distinguishes positive, hard-negative and easy-negative labels. Training-label precision reports exact-category matches among the 87 evaluation pairs labelled positive or hard negative by the LLM. Error bars indicate 95\% confidence intervals from 10,000 bootstrap resamples stratified by dataset. The 95\% confidence intervals were 61.3--76.0\% for three-class agreement, 86.7--95.3\% for binary agreement and 74.7--90.2\% for training-label precision. Four pairs with tied highest three-class vote counts were assigned hard-negative reference labels.}
\label{fig:results_overall}
\end{figure}

We evaluated SYNTRA across 24 retrieval datasets spanning 15 domains using the same synthesis workflow (Fig.~\ref{fig:results_overall}a). Training data were generated by Qwen3-30B-A3B \cite{yang2025qwen3}, deployed on four NVIDIA GeForce RTX 5090 GPUs. We adapted a Qwen3-0.6B backbone separately for each task using low-rank adaptation (LoRA) \cite{hu2022lora} and compared the resulting models with an MS MARCO-trained baseline \cite{nguyen2016msmarco} using the same backbone and training procedure. Retrieval performance was measured by nDCG@10 on a 0--100 scale.

Models trained on SYNTRA-generated data outperformed the baseline in every task-specific comparison shown in Fig.~\ref{fig:results_overall}a. Gains spanned specialised knowledge domains and tasks involving mathematical, spatial and commonsense reasoning. Relative nDCG@10 improvements were 32.4\% on LeCaRDv2 and 36.0\% on SCIDOCS, representing legal and scientific retrieval, respectively. Gains reached 94.5\% on TheoremQA questions and 97.1\% on WinoGrande, with the largest relative improvement observed on SpartQA, at 416.6\%. Although their magnitude varied, the consistently positive gains demonstrate that the same synthesis workflow can produce effective supervision across diverse corpora and task-specific information needs.

Human involvement was confined to task setup, typically requiring no more than 10\,min per task according to retrospective estimates. Subsequent synthesis-instruction refinement and data production proceeded automatically. Selected task-level production times are reported in Extended Data Table~\ref{tab:ed-selected-production-runtime}, and implementation and training details are provided in \hyperref[sec:methods]{Methods}.

\subsection*{Self-improvement increases the training value of synthetic data}\label{subsec:self-refinement}

Automatic self-improvement increased the training value of SYNTRA-generated data on all six retrieval tasks examined (Fig.~\ref{fig:results_overall}b). The largest relative gains occurred on NFCorpus and FiQA, where refining synthesis instructions improved downstream nDCG@10 by 45.0\% and 23.1\%, respectively, compared with synthesis using the initial instructions. On both tasks, instruction refinement changed the outcome of adaptation: models trained on data generated with the initial instructions underperformed the MS MARCO-trained baseline, whereas those trained on data generated with the refined instructions surpassed it. The benefits of instruction refinement extended to TheoremQA theorems, ArguAna, SCIDOCS and Text2SQL, with relative gains of up to 14.2\%.

Automatic self-improvement provided an effective alternative to human calibration of synthesis instructions, yielding higher downstream retrieval scores on four of the six tasks and comparable performance on Text2SQL (Fig.~\ref{fig:results_overall}b). Human calibration remained more effective on TheoremQA theorems. This gap may reflect the specialised mathematical judgement needed to distinguish relevant theorems from superficially related ones, which sparse examples and automated feedback may not fully capture. The results suggest that automatic refinement can handle task-specific synthesis design across diverse settings, while expert calibration may remain valuable when relevance depends on specialised reasoning.

\begin{figure}[!htbp]
\centering
% Remove blank outer canvas while preserving the current manuscript width.
\includegraphics[width=130.7mm,trim=0 2mm 0 2mm,clip]{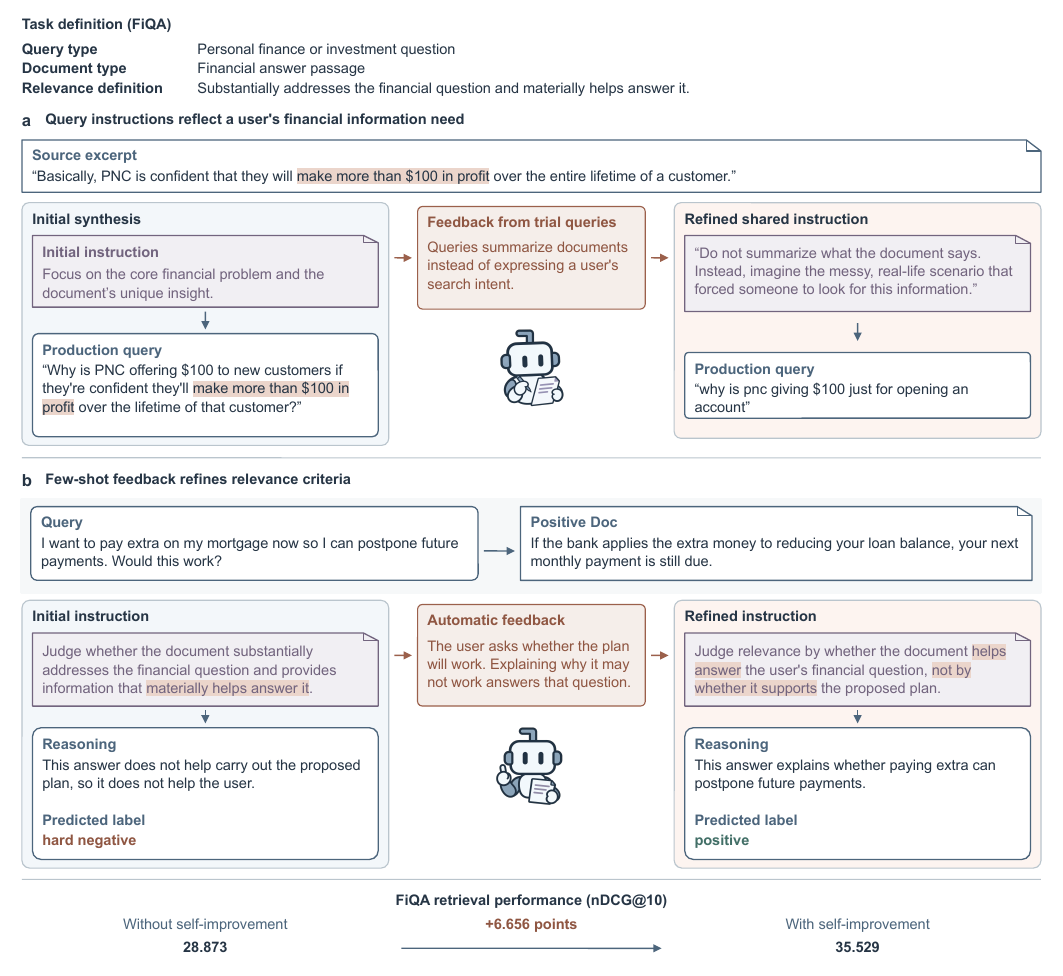}
\caption{\textbf{Few-shot-guided self-improvement on FiQA.}
\textbf{a}, Refinement of query-generation instructions. The source excerpt and queries are verbatim production examples generated in separate runs with instructions corresponding to the first and third refinement rounds. Attribute plans were not held fixed between these runs. Feedback summarizes the original refinement history.
\textbf{b}, Refinement of relevance-annotation instructions using a known-positive few-shot example. Feedback clarifies that explaining why a plan may fail can help answer the user's question. Text in this panel provides an illustrative summary of the recorded calibration episode.
Bottom, FiQA retrieval performance without and with self-improvement, reflecting joint refinement of query-generation and relevance-annotation instructions.}
\label{fig:self-improvement-case}
\end{figure}

A case study on FiQA, a financial question-answering retrieval task, illustrates how feedback translated task requirements into reusable synthesis instructions (Fig.~\ref{fig:self-improvement-case}). For query generation, feedback revealed that trial queries summarised source documents rather than expressed the financial questions users might ask. The revised instructions redirected synthesis towards the user's search intent (Fig.~\ref{fig:self-improvement-case}a). For relevance assessment, an example provided as a relevant query-document pair exposed an overly narrow interpretation of helpfulness. The query asked whether paying extra on a mortgage would allow future payments to be postponed. Although the document explained why this would not work, the annotator classified it as a hard negative because it did not help carry out the proposed plan. Feedback prompted revised instructions clarifying that relevance depends on helping answer the query, rather than supporting the proposed plan. The same example was then correctly labelled positive (Fig.~\ref{fig:self-improvement-case}b). In both cases, the revisions updated instructions reused across documents, allowing feedback on individual trial outputs to guide subsequent data synthesis. Complete instructions and example inputs are provided in Supplementary Note 1.

\begin{figure}[!htbp]
\centering
\includegraphics[width=\textwidth]{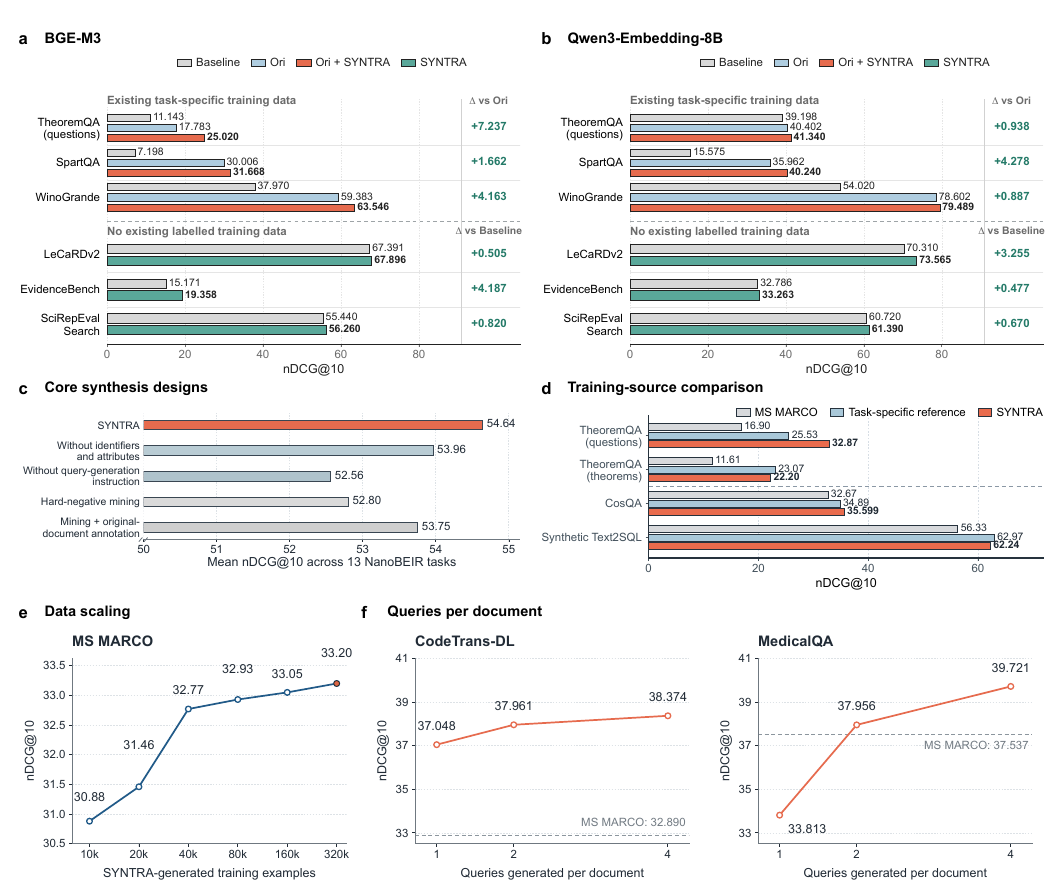}
\caption{\textbf{Model adaptation and ablations of SYNTRA.}
\textbf{a,b}, BGE-M3 (\textbf{a}) and Qwen3-Embedding-8B (\textbf{b}) results with and without existing task-specific labelled training data. Ori denotes models fine-tuned on existing task-specific labelled data. Dashed lines separate the two settings; right-hand values report gains over Ori or the corresponding pretrained baseline.
\textbf{c}, NanoBEIR ablations of the core synthesis components.
\textbf{d}, Comparison of the shared MS MARCO baseline, a task-specific reference and SYNTRA. Reference training data were constructed using ReasonEmbed queries for the two TheoremQA tasks and official queries for CosQA and Synthetic Text2SQL.
\textbf{e}, MS MARCO nDCG@10 as SYNTRA-generated training data increase from 10,000 to 320,000 examples.
\textbf{f}, Retrieval performance as the number of queries generated per document increases on CodeTrans-DL and MedicalQA. Dashed lines show the MS MARCO-trained reference scores.}
\label{fig:results_analysis}
\end{figure}

\subsection*{SYNTRA-generated queries outperform human search queries and rival task-specific alternatives}\label{subsec:rq1}

We compared SYNTRA-generated queries with human queries from Bing Search logs to assess their value as retrieval supervision. Both query sources were associated with the same pool of 10,000 MS MARCO source documents and underwent the same relevance-annotation and negative-mining procedures. The resulting datasets were used to train the same embedding backbone with an identical training recipe.

Replacing human search queries with SYNTRA-generated queries increased mean nDCG@10 from 42.18 to 42.87 on BEIR and from 54.53 to 55.51 on NanoBEIR, corresponding to relative improvements of 1.6\% and 1.8\%, respectively (Fig.~\ref{fig:results_overall}c). These gains show that document-grounded query synthesis can even provide more effective retrieval supervision than human search queries when the same training setup is used.

We further evaluated queries generated using self-improved synthesis instructions against task-specific query sources: queries released by ReasonEmbed for TheoremQA questions and theorems \cite{chen2026reasonembed}, and official queries for CosQA \cite{huang2021cosqa} and Synthetic Text2SQL. Models trained with SYNTRA-generated queries outperformed those trained with task-specific queries on TheoremQA questions and CosQA, while remaining competitive on TheoremQA theorems and Synthetic Text2SQL (Fig.~\ref{fig:results_analysis}d). These results show that a single synthesis workflow, adapted to each task through self-improvement, can rival task-specific query sources without requiring manual redesign of query-generation instructions.

\subsection*{SYNTRA-annotated relevance labels align with human judgements}\label{subsec:rq2}

We evaluated relevance labels annotated by Qwen3-30B-A3B against reference labels derived from votes by seven independent human annotators. The evaluation covered 150 query-document pairs, with 50 pairs each from MS MARCO, Text2SQL and TheoremQA (theorems). Dataset-level agreement and class-wise performance are reported in Extended Data Tables~\ref{tab:rq2-agreement-summary} and \ref{tab:rq2-confusion}.

Agreement with human judgements was 91.3\% for binary relevance classification (137/150; Cohen's $\kappa = 0.719$) and 68.7\% for three-class classification into positives, hard negatives and easy negatives (103/150; $\kappa = 0.510$). The lower three-class agreement largely reflected differences in negative-difficulty assessment: 34 of the 47 disagreements involved hard versus easy negatives rather than relevant versus non-relevant documents (Extended Data Table~\ref{tab:rq2-confusion}). We further examined the positive and hard-negative labels retained for training. Among the 87 pairs assigned either label by the LLM, 72 matched the human reference category exactly, yielding a pooled precision of 82.8\% (Fig.~\ref{fig:results_overall}d; Extended Data Table~\ref{tab:rq2-confusion}).

Human inter-annotator agreement, measured by Fleiss' $\kappa$, was 0.483 for three-class labels and 0.606 for binary labels. After excluding the four pairs with tied highest vote counts, three-class agreement between LLM-generated labels and human reference labels was 69.2\% (101/146; Cohen's $\kappa = 0.520$), compared with 68.7\% (103/150; $\kappa = 0.510$) for all pairs. Thus, the treatment of tied votes had little effect on the overall agreement.

Together, these findings support SYNTRA's relevance annotation as a source of retrieval supervision, including the positive and hard-negative labels used for training. Disagreements were concentrated in the distinction between hard and easy negatives rather than binary relevance.

\subsection*{Strong embedding models benefit from SYNTRA-generated training data}\label{subsec:rq45}

The cross-task experiments used a controlled backbone, leaving open whether SYNTRA-generated data also benefit strong embedding models already trained on large curated corpora. We adapted BGE-M3 \cite{chen2024m3} and Qwen3-Embedding-8B \cite{zhang2025qwen3embedding} in two practical settings: tasks without existing labelled training sets (LeCaRDv2, EvidenceBench and SciRepEval Search) and tasks with such sets (TheoremQA questions, SpartQA and WinoGrande) (Fig.~\ref{fig:results_analysis}a,b).

For tasks without existing labelled training sets, SYNTRA generated training data guided by a small set of relevant query-document examples. Fine-tuning on these data improved both embedding models on all three tasks compared with their performance before task-specific fine-tuning. On LeCaRDv2, EvidenceBench and SciRepEval Search Tasks, BGE-M3 achieved relative nDCG@10 improvements of 0.7\%, 27.6\% and 1.5\%, respectively. The corresponding gains for Qwen3-Embedding-8B were 4.6\%, 1.5\% and 1.1\%.

For tasks with existing labelled training sets, we compared fine-tuning on those sets alone with fine-tuning on the same sets augmented by SYNTRA-generated data. Augmentation improved both models on all three tasks. On TheoremQA questions, SpartQA and WinoGrande, BGE-M3 achieved relative nDCG@10 gains of 40.7\%, 5.5\% and 7.0\%, respectively, over fine-tuning on labelled data alone. The corresponding gains for Qwen3-Embedding-8B were 2.3\%, 11.9\% and 1.1\%.

Together, these results show that SYNTRA-generated data benefit strong embedding models both by supplying task-specific supervision where labelled training sets are unavailable and by complementing existing supervision where they are available.

\subsection*{Scaling synthetic data yields continued retrieval gains}\label{subsec:rq6}

We trained a Qwen3-8B backbone on SYNTRA-generated datasets constructed from MS MARCO passages, doubling the dataset size from 10,000 to 320,000 examples across six scales. The same training recipe and MS MARCO evaluation protocol were used throughout.

Retrieval performance improved at every evaluated scale (Fig.~\ref{fig:results_analysis}e). Across the 32-fold expansion, nDCG@10 increased from 30.88 to 33.20, corresponding to a relative gain of 7.5\%. These results show that SYNTRA-generated data retain their training value as production scales, with larger datasets yielding continued retrieval gains.

\subsection*{Core design choices improve retrieval performance}\label{subsec:rq7}

To identify contributions from SYNTRA's synthesis components, we evaluated task-adapted query instructions, planning with semantic identifiers and query attributes, relevance annotation, and multiple queries per document. For the component ablations, we report mean nDCG@10 across 13 NanoBEIR tasks, a lower-cost proxy that preserved the relative BEIR performance of 18 public embedding models (Methods). Query-count comparisons are reported separately for each evaluated task (Fig.~\ref{fig:results_analysis}c,f).

Relative to variants without these components, the task-adapted query-generation instruction improved mean nDCG@10 on NanoBEIR by 4.0\%, and planning with semantic identifiers and query attributes improved it by 1.3\%. For relevance annotation, the full workflow improved mean nDCG@10 by 3.5\% over hard-negative mining alone. Adding relevance annotation for the source document to hard-negative mining yielded a 1.8\% gain, but the full workflow remained 1.7\% higher. These comparisons support the contributions of task-adapted query instructions, semantic identifiers and query attributes, and relevance annotation to the effectiveness of the generated training data.

Generating multiple queries per document also improved retrieval on both evaluated tasks (Fig.~\ref{fig:results_analysis}f). Different combinations of semantic identifiers and query attributes allow the same document to support queries expressing different information needs. Compared with one query per document, generating two and four queries increased nDCG@10 by 2.5\% and 3.6\% on CodeTrans-DL and by 12.3\% and 17.5\% on MedicalQA, respectively. These results support generating several queries per document as a way to provide additional useful supervision on these tasks.

% ============================================================================
\FloatBarrier
\section{Discussion}\label{sec:discussion}

SYNTRA demonstrates that self-improvement can turn guidance from a small set of retrieval examples into reusable instructions for task-specific data synthesis. The examples convey what users seek and what makes a document relevant, while SYNTRA uses feedback on trial outputs to iteratively refine instructions for generating queries and assessing candidate documents. Because these revisions are incorporated into instructions reused across documents, their influence extends beyond the examples examined during refinement. This provides a means of automating part of the task-specific data design needed to adapt embedding models across domains and tasks.

Self-improvement is effective because instructions inferred from a few examples can miss requirements that only become apparent in trial outputs. In the FiQA case study, trial queries summarised source documents rather than asking the financial questions users would pose, and the relevance instruction rejected an answer that explained why a proposed plan would fail, even though such an answer resolves the user's question. Correcting such mismatches substantially affected downstream learning: on NFCorpus and FiQA, data generated with the initial instructions trained models that underperformed the MS MARCO-trained baseline, whereas data generated after refinement surpassed it. Automatic refinement outperformed human calibration on four of six tasks and matched it on a fifth, suggesting that many such mismatches can be detected without expert input. On TheoremQA theorems, however, human calibration remained more effective, possibly because distinguishing relevant theorems from superficially related ones requires mathematical judgement that a few examples and automated feedback do not fully capture.

Separating instruction refinement from data production also makes this approach practical to scale. Commercial-model calls are confined to small-sample testing, after which the refined instructions are reused for large-scale generation with a locally deployed model; production runs for most tasks took 2--11\,h on four consumer GPUs (Extended Data Table~\ref{tab:ed-selected-production-runtime}). The same synthesis workflow can therefore support different tasks without manual instruction refinement for each one, with human involvement limited to task setup. Within the retrieval settings examined here, the resulting data improved both a controlled backbone and strong embedding models, and retained their training value as data volume increased. More advanced strategies for optimizing language-model prompts and pipelines \cite{yang2024opro,yuksekgonul2025textgrad} could be incorporated into instruction refinement to further improve the synthesized data.

These findings motivate extending self-improvement to data synthesis for other LLM capabilities, building on recent work on automated question-answer construction with assessment criteria and task construction from teacher demonstrations \cite{li2026ares,zhao2026nexforge}. Such extensions would require feedback suited to the target capability, such as answer checks and code execution for mathematical and coding tasks \cite{guo2025deepseekr1} or task-specific rubrics for relevance, completeness and contextual requirements \cite{gunjal2026rubrics}. Feedback from model evaluation, task execution and user interactions could further reveal unmet or emerging training needs. Connecting these signals to instruction refinement could extend self-improvement towards identifying what models need to learn next and adapting data synthesis accordingly, both for the retrieval components of AI agents and for LLM training more broadly.

% ============================================================================
\FloatBarrier
\section{Methods}\label{sec:methods}

We first describe corpus preprocessing and task preparation, followed by the synthesis workflow and the self-improvement procedure used to adapt its instructions. We then detail embedding-model training, experimental design and evaluation, and human evaluation of relevance labels.

\subsection{Corpus preprocessing and task preparation}\label{subsec:task-preparation}

SYNTRA takes an unlabelled corpus and a small set of examples pairing queries with relevant documents (typically \textbf{five} per task in our experiments) (Fig.~\ref{fig:syntra-overview}a). In our experiments, these examples were randomly sampled from the benchmark's public release on Hugging Face, which for some tasks is the evaluation split. These pairs served only as in-context demonstrations for task-definition inference, instruction refinement and query planning; they were not added to the training data. To support candidate retrieval, document titles and bodies are concatenated where both are available, encoded as L2-normalised vectors and stored in a full-corpus FAISS inner-product index \cite{douze2024faiss}.

A separate, diversity-aware selection procedure identifies source documents for query generation (Algorithm~\ref{alg:document-selection}). Using an auxiliary inverted-file (IVF) index, the procedure processes documents in shuffled order. After selecting a document, it excludes neighbouring documents whose similarity to that document exceeds the selection threshold, reducing semantic redundancy in the selected set. Selection continues until the target count is reached or the available pool is exhausted. Where feasible, we selected 10,000 source documents. Test mode uses small samples from this set, whereas production mode processes the selected set using the final synthesis instructions. Document selection leaves the full-corpus index intact, allowing queries to retrieve candidates from the entire corpus in both modes.

\begin{algorithm}[t]
\caption{Corpus indexing and diversity-aware document selection}
\label{alg:document-selection}
\begin{algorithmic}[1]
\Require Corpus $\mathcal{C}$, encoder $f$, target count $M$, similarity threshold $\tau$
\Ensure Selected documents $S$ and full-corpus retrieval index $J$
\State Encode each document $d$ as $\mathbf{v}_d=f(d)/\lVert f(d)\rVert_2$
\State Build inner-product index $J$ over all document vectors
\State Build an auxiliary IVF index $J_{\mathrm{sel}}$ for neighbour search
\State $L\gets\Call{Shuffle}{\mathcal{C}}$; $A\gets\mathcal{C}$; $S\gets\varnothing$
\For{each document $d$ in $L$}
    \If{$|S|=M$ or $A=\varnothing$}
        \State \textbf{break}
    \EndIf
    \If{$d\in A$}
        \State $S\gets S\cup\{d\}$; $A\gets A\setminus\{d\}$
        \State $N_d\gets\Call{AvailableNeighbours}{J_{\mathrm{sel}},\mathbf{v}_d,A}$
        \State $A\gets A\setminus\{d'\in N_d:\mathbf{v}_d^{\top}\mathbf{v}_{d'}>\tau\}$
    \EndIf
\EndFor
\State \Return $S,J$
\end{algorithmic}
\end{algorithm}

Task preparation establishes the requirements that guide data synthesis. From the supplied query-document examples, SYNTRA infers a task definition specifying the intended query type, document type and relevance criteria. A generate-evaluate-refine procedure checks the proposed definition against the examples and revises requirements that are inaccurate or insufficiently specific. Alternatively, users can provide a task definition directly. The resulting definition guides the construction and subsequent refinement of synthesis instructions in test mode, as described below.

Human involvement in the principal experiments was limited to task setup. The operator reviewed and occasionally revised the inferred task definition and, when necessary, adjusted the document-selection threshold to obtain the intended number of source documents. Based on retrospective estimates, these activities typically required no more than 10\,min per task.

\subsection{Self-improving agentic framework for retrieval data synthesis}

SYNTRA constructs retrieval training data through identifier extraction and attribute construction, query synthesis, and relevance annotation (Fig.~\ref{fig:syntra-production}). These operations form a common synthesis workflow used in both \emph{test mode} and \emph{production mode}. In test mode, feedback on trial outputs guides revisions to the synthesis instructions. In production mode, the final instructions remain fixed and are reused across the selected documents. We first describe the synthesis operations, followed by the self-improvement procedure that adapts their instructions to the target task.

\subsubsection{Identifier extraction and attribute construction}\label{subsec:identifier-attribute-construction}

To support queries expressing different information needs, SYNTRA identifies what can be asked about a document and how those questions can vary. For each selected document, a single LLM call jointly extracts semantic identifiers and constructs query attributes, using instructions conditioned on the task definition. Semantic identifiers are document-grounded entities or relations that provide possible targets for an information need. Query attributes specify dimensions of variation, such as scope, reasoning requirements or specificity, together with task-appropriate options for each dimension. Both the identifiers and the attribute dimensions and options are constructed for each document. Their combinations provide the basis for planning distinct queries (Fig.~\ref{fig:syntra-production}a).

\subsubsection{Query synthesis}\label{subsec:query-synthesis}

Query synthesis separates planning an information need from expressing it as a query (Fig.~\ref{fig:syntra-production}b). The planning stage receives the source document, its semantic identifiers and query attributes, the task definition, the supplied query-document examples and the task-adapted query instruction. Each plan combines one identifier with one option from every attribute dimension to specify an information need and its intended expression. The query-writing stage then converts each plan into a query using the source document and the same task-adapted instruction. Generating multiple plans allows a document to support queries with different information needs or forms of expression.

\subsubsection{Relevance annotation}\label{subsec:label-synthesis}

Each generated query retrieves its top-$K$ candidate documents from the full-corpus index using the indexing encoder and cosine similarity. SYNTRA assesses these candidates against the task-specific relevance criteria (Fig.~\ref{fig:syntra-production}c). The source document is assessed in the same way if it is retrieved; its use for query generation does not automatically make it a positive.

Candidates are processed in chunks, with each query-document pair judged independently on a four-point scale. A score of 3 indicates strong satisfaction of the relevance criteria, whereas 2 indicates a weaker but valid positive. A score of 1 denotes a plausible candidate that fails the relevance criteria, and 0 denotes a clearly unrelated candidate. Scores of 2 and 3 are mapped to positives, 1 to hard negatives and 0 to easy negatives. Annotations from all chunks are combined into the corresponding document sets for each query. These outputs are converted into training records as described below.

\subsubsection{Instruction self-improvement}\label{subsec:task-adaptation}

In \emph{test mode}, SYNTRA adapts the instructions governing these operations through generate-evaluate-refine cycles on small samples (Fig.~\ref{fig:syntra-overview}b). Trial outputs are evaluated against the task definition and the supplied query-document examples. Diagnostic feedback guides instruction revisions, after which synthesis is repeated on the same trial inputs. The revisions apply to instructions reused across documents, allowing feedback from a small sample to guide subsequent data production.

Three refinement loops address identifier extraction and attribute construction, query synthesis, and relevance annotation. Each loop allows up to three attempts by default, including the initial trial; this limit is configurable. For the identifier-attribute and query loops, LLM verifiers assign scores from 0 to 3, with scores of at least 2 treated as a pass. Relevance annotation uses the programmatic consistency check described below. Each loop ends when verification passes or the attempt limit is reached. SYNTRA retains the last evaluated instruction, including when verification has not passed at the limit.

\paragraph{Identifier and attribute refinement.}
Identifier extraction and attribute construction are refined together on sampled documents. The initial trial uses a generation template conditioned on the task definition, without additional refinement guidance. An LLM verifier assesses whether the outputs are grounded in the document, whether they contain redundant elements and whether the attribute dimensions and options support useful variation within the task. Feedback is used to develop reusable guidance for subsequent extraction and construction. Once this loop finishes, identifiers and attributes are regenerated using the final guidance and held fixed during query-instruction refinement.

\paragraph{Query-instruction refinement.}
The initial query instruction is derived from the task definition and the supplied examples. It specifies the intended query producer and information need, how document evidence should be selected or abstracted, and the linguistic characteristics and constraints of the target queries. SYNTRA applies this instruction to both information-need planning and query writing. An LLM verifier evaluates sampled query-document pairs for task alignment, style and naturalness, including whether the queries plausibly reflect their intended producer. It also assesses the effectiveness of the instruction. Feedback guides revisions to the instruction used in both stages, while the identifiers and attributes remain unchanged.

\paragraph{Relevance-instruction refinement.}
The initial annotation instruction is derived from the task definition and the required annotation format. SYNTRA tests it on the supplied query-document pairs, each of which contains a document known to be relevant to its query. A program checks whether every pair is classified as positive, treating missing predictions as mismatches. If the check fails, the LLM analyses the errors and revises the relevance criteria. It also develops criteria for distinguishing plausible candidates that fail a required condition from documents unrelated to the information need. The supplied examples directly verify positive classifications; criteria for negative categories are inferred from the task requirements. The framework also supports calibration against human-confirmed labels during test mode.

All refinement loops finish before \emph{production mode} begins. The final instructions are then held fixed and applied across the selected documents using the synthesis operations described above. Identifiers, attributes, queries and relevance labels continue to be generated for individual documents and queries. Parsing retries and structural checks are applied to production outputs.

\subsubsection{Computational resources and production times}

Instruction refinement in test mode typically used Gemini 3 Flash Preview (Supplementary Note 2). Data generation used Qwen3-30B-A3B deployed on four NVIDIA GeForce RTX 5090 GPUs within a six-GPU system. The remaining two GPUs were used for embedding-model operations. Selected production batches across the 24 retrieval tasks are reported in Extended Data Table~\ref{tab:ed-selected-production-runtime}. Reported runtimes cover data production and exclude calibration, prebuilt indexing, and embedding-model training and evaluation. The table caption details how batches were selected and which stages were included in the timings.

\subsection{Embedding-model training}\label{subsec:training-evaluation}

\paragraph{Training-data preparation.}
We converted the synthesized queries and annotated documents into training records, retaining queries with at least one positive and one usable negative. Each record contained the query, lists of positive and negative document texts, and a retrieval instruction that could be prepended to the query during training. Training-data exports supported two negative-selection policies: using only annotated hard negatives, or supplementing them with annotated easy negatives when fewer than the requested number of hard negatives were available.

\paragraph{Models and training configurations.}

Embedding-model training began after data synthesis was complete. The main cross-task experiments used a Qwen3-0.6B backbone, adapted separately for each task using low-rank adaptation (LoRA) \cite{hu2022lora}. We also fine-tuned BGE-M3 \cite{chen2024m3} and Qwen3-Embedding-8B \cite{zhang2025qwen3embedding} to assess the value of synthetic data for strong pretrained embedding models. To evaluate the effect of increasing the volume of synthetic training data, we used a Qwen3-8B backbone.

For the Qwen3-based models, the training data loader sampled one positive and a configurable number of negatives from each record. When too few negatives were available, their indices were repeated to supply the requested number. Model-specific training settings are provided in Supplementary Note 2 (Supplementary Tables 1--2). Experimental comparisons and evaluation procedures are described below.

\subsection{Experimental design and evaluation}\label{subsec:experimental-evaluation}

We evaluated the training value of SYNTRA-generated data across retrieval tasks, embedding models, query sources and data scales, and used ablations to assess the contributions of self-improvement and individual synthesis components. Retrieval performance was measured by normalised discounted cumulative gain at rank 10 (nDCG@10), reported on a 0--100 scale. Training-data preparation and model configurations are described above, with detailed settings in Supplementary Note 2 (Supplementary Tables 1--2). Human evaluation of relevance labels is described separately below.

\subsubsection{Retrieval adaptation across tasks and models}

The main cross-task evaluation covered 24 retrieval datasets spanning 15 domains (Fig.~\ref{fig:results_overall}a). SYNTRA generated training data separately for each task using the same synthesis workflow. Models trained on these data were compared with an MS MARCO-trained baseline \cite{nguyen2016msmarco}, using the same Qwen3-0.6B backbone and training procedure. Each task-specific model was evaluated on its corresponding retrieval task. MS MARCO provided the common training reference, while the remaining 23 datasets formed the task-adaptation comparisons.

We also evaluated BGE-M3 and Qwen3-Embedding-8B in two settings (Fig.~\ref{fig:results_analysis}a,b). For tasks without existing labelled training sets---LeCaRDv2, EvidenceBench and SciRepEval Search \cite{singh2023scirepeval}---models fine-tuned on SYNTRA-generated data were compared with their pretrained counterparts without task-specific fine-tuning. For tasks with existing labelled training sets---TheoremQA questions \cite{su2025bright}, SpartQA \cite{mirzaee2021spartqa} and WinoGrande \cite{sakaguchi2020winogrande}---we compared fine-tuning on those sets alone with fine-tuning on the same sets augmented by SYNTRA-generated data.

\subsubsection{Query-source comparisons}

\paragraph{Human search queries.}
The MS MARCO experiment compared human search queries from Bing logs \cite{nguyen2016msmarco} with SYNTRA-generated queries using the same pool of 10,000 source documents. Both query sources underwent the same relevance-annotation and negative-mining procedures. The resulting datasets were used to train the same embedding backbone with the same training recipe, and retrieval was evaluated on BEIR and NanoBEIR (Fig.~\ref{fig:results_overall}c). Valid training-record counts are provided in Extended Data Table~\ref{tab:ed-selected-production-runtime}.

\paragraph{Task-specific query sources.}
We compared SYNTRA-generated queries with task-specific sources on TheoremQA questions and theorems, CosQA and Synthetic Text2SQL (Fig.~\ref{fig:results_analysis}d). Reference queries were those released by ReasonEmbed for the two TheoremQA tasks \cite{chen2026reasonembed} and the official training queries for CosQA \cite{huang2021cosqa} and Synthetic Text2SQL. Each reference query was paired with its released positive document, which served as its source document. Within each task, both query sources then underwent the same candidate-retrieval and relevance-annotation procedures and were used to train the same embedding backbone, so that the comparison differed only in the source of the queries.

For these four tasks, we retained records containing at least one annotated positive and one hard negative. From each retained record, we constructed two positive-selection variants. The first retained all annotated positives. The second retained only the source document if it had been labelled positive; otherwise, it retained the original positive list. Both variants used the same hard negatives and were concatenated without deduplication for training. The combined number of input rows was therefore twice the number of retained records. Query counts, records per variant and combined input rows are reported in Extended Data Table~\ref{tab:query-source-comparisons}.

\subsubsection{Self-improvement and component ablations}

\paragraph{Self-improvement.}
We compared three conditions across six retrieval tasks: TheoremQA theorems, ArguAna, NFCorpus, FiQA, SCIDOCS and Text2SQL (Fig.~\ref{fig:results_overall}b). Synthesis without self-improvement used the initial task-specific instructions for production. Automatic self-improvement used instructions revised through feedback in test mode, following the instruction self-improvement procedure described above. In the human-calibration condition, human input was used to calibrate synthesis instructions during test mode. Data generated under each condition were used to train task-specific embedding models. Their retrieval performance is reported in Fig.~\ref{fig:results_overall}b.

\paragraph{Synthesis components.}
We evaluated the contributions of task-adapted query instructions, semantic identifiers and query attributes, and relevance annotation (Fig.~\ref{fig:results_analysis}c). For query synthesis, we compared the full workflow with variants that omitted either the task-adapted query instruction or semantic identifiers and query attributes. For relevance annotation, we compared the full workflow with hard-negative mining alone and with hard-negative mining supplemented by source-document annotation. Component ablations used a separate sample of 10,000 MS MARCO source documents from that used in the query-source comparison; all variants, including the full workflow (54.64), were generated from this same document pool. To reduce the cost of evaluating multiple synthesis variants, we used NanoBEIR and reported mean nDCG@10 across its 13 tasks. We also assessed the consistency of model comparisons between NanoBEIR and BEIR by evaluating 18 publicly available embedding models on both suites, as described below.

\paragraph{Queries per document.}
We varied the number of queries generated per document from one to two and four on CodeTrans-DL and MedicalQA (Fig.~\ref{fig:results_analysis}f). SYNTRA used different combinations of semantic identifiers and query attributes to generate queries expressing different information needs. Retrieval performance was reported separately for each task. This comparison assessed the benefit of generating additional queries per document; the resulting training datasets therefore differed in both query count per document and total size.

\subsubsection{Scaling synthetic training data}

We evaluated the effect of increasing synthetic training-data volume using a Qwen3-8B backbone. Starting with 10,000 SYNTRA-generated examples from MS MARCO passages, we increased the dataset size to 20,000, 40,000, 80,000, 160,000 and 320,000 examples. The training recipe and MS MARCO evaluation protocol were kept unchanged across scales (Fig.~\ref{fig:results_analysis}e). We performed one training run at each scale using the settings in Supplementary Note 2 (Supplementary Table 2).

\subsubsection{Consistency between NanoBEIR and BEIR}
\label{sec:nanobeir-beir-consistency}

We assessed whether NanoBEIR preserved the relative model performance observed on BEIR \cite{thakur2021beir} by evaluating 18 publicly available embedding models on both suites. The analysis included nine encoder-only and nine decoder-only models. For each model, we obtained paired suite-level mean nDCG@10 scores; the models and their scores are listed in Extended Data Table~\ref{tab:nanobeir-beir-models}.

NanoBEIR evaluation used the MTEB implementation \cite{muennighoff2023mteb} and comprised 13 retrieval tasks: ArguAna, Climate-FEVER, DBpedia, FEVER, FiQA2018, HotpotQA, MS MARCO, NFCorpus, Natural Questions, Quora, SCIDOCS, SciFact and Touche2020. Each task contained 50 evaluation queries except Touche2020, which contained 49, giving 649 queries in total. We first averaged nDCG@10 over the evaluation queries within each task, then calculated the unweighted arithmetic mean across the 13 tasks and multiplied it by 100. Each task therefore contributed equally to the suite score.

To assess agreement in model ordering without conflating within-family rankings with differences between architecture families, we analysed encoder-only and decoder-only models separately. Within each family, we calculated Spearman's rank correlation between the paired BEIR and NanoBEIR scores, with two-sided significance tests. We also calculated each model's absolute change in rank between the suites, summarised these changes by their median, and counted models whose ranks differed by no more than one position. NanoBEIR closely preserved the BEIR ordering within both families (encoder-only: Spearman's $\rho=0.933$, $P=2.0\times10^{-4}$; decoder-only: $\rho=0.817$, $P=7.0\times10^{-3}$). The median absolute rank change was one position in each family, and 8 of 9 encoder-only and 6 of 9 decoder-only models differed by no more than one rank (Extended Data Tables~\ref{tab:nanobeir-beir-models} and \ref{tab:nanobeir-beir-summary}). These results support the use of NanoBEIR as a lower-cost proxy for BEIR in the component ablations.

\subsection{Human evaluation of relevance labels}\label{subsec:human-label-evaluation}

Seven annotators independently assessed 150 query-document pairs, comprising 50 pairs from each of MS MARCO, Text2SQL and TheoremQA theorems. All annotators had received undergraduate or graduate training in advanced mathematics and computer science. Annotators used the same four-point relevance scale as SYNTRA's relevance annotation, with ratings mapped to positives, hard negatives and easy negatives. Three-class human reference labels were determined by plurality voting, with ties assigned to the hard-negative category. For binary evaluation, each annotator's hard-negative and easy-negative labels were merged into a single non-relevant category before majority voting.

We measured agreement between SYNTRA-annotated labels and human reference labels using exact agreement and Cohen's $\kappa$. Human inter-annotator agreement was measured using Fleiss' $\kappa$ for the three-class and binary labels. Pooled training-label precision was defined as the proportion of pairs labelled positive or hard negative by the LLM whose predicted category exactly matched the human reference.

We estimated 95\% percentile confidence intervals using 10,000 bootstrap resamples, each comprising 50 pairs sampled with replacement from each dataset. Human reference labels were held fixed, and training-label precision was recalculated for each resample. To assess sensitivity to tie handling, we repeated the three-class LLM-human comparison after excluding pairs with tied highest vote counts.

\paragraph{Ethics.}
The seven annotators were volunteers: three were authors of this study and the others were graduate students in the authors' research group. They rated pairs consisting of SYNTRA-generated queries and documents from publicly available benchmark corpora. No personal information about the annotators was collected, and no data about the annotators themselves were analysed. Their ratings served only as reference labels for evaluating SYNTRA-annotated relevance labels. All annotators participated voluntarily without compensation and gave verbal informed consent after being informed of the purpose of the study. Because the annotators were not themselves the objects of testing, surveys or observation, and no human biological samples or personal information data were used, the study did not constitute research involving human participants as defined in Article 2 of China's Measures for Science and Technology Ethics Review (Trial) (2023) and therefore did not require ethics review.

% ============================================================================
\backmatter
% ============================================================================

\bmhead{Data availability}

SYNTRA retrieval-training data exports, including synthetic-query data, original-query controls and archived experimental variants, are publicly available at \url{https://huggingface.co/datasets/ZucanLyu/SYNTRA}. The repository provides file manifests and describes the audited exports. Source benchmark data are available from their original providers, as cited in this Article, and remain subject to their respective access conditions and terms.

\bmhead{Code availability}

The SYNTRA source code is publicly available at \url{https://github.com/zucanlv/SYNTRA}. The repository includes the data synthesis pipeline, training and retrieval evaluation scripts, code for human relevance evaluation, configuration files, and installation and workflow documentation.

% Keep each reference entry together, including its final URL lines.
\begingroup
\interlinepenalty=10000
\bibliography{sn-bibliography}
\endgroup

\bmhead{Funding}

The work was supported by grants from National Natural Science Foundation of China (No. U24A20253) and Scientific Research Innovation Capability Support Project for Young Faculty.

\bmhead{Acknowledgements}

We thank the annotators for their contributions to the human relevance annotation study. We also thank Zheng Liu for valuable suggestions on this work.

\bmhead{Author contributions}

Z.L., J.C. and D.L. conceived the study and designed the methodology. Z.L. developed the software, prepared the corpora and retrieval tasks, and produced the synthetic data. Z.L. and J.C. designed the experiments, and Z.L. and R.T. performed them. Z.L. performed the human calibration of synthesis instructions. Z.L. and J.C. designed the human annotation study, and Z.L., R.T. and P.J. served as annotators. Z.L. and R.T. analysed the results and prepared the figures and tables. Z.L., R.T., J.C. and D.L. wrote and revised the manuscript. J.C. and D.L. supervised the research and provided resources.

\bmhead{Competing interests}
\enlargethispage{2\baselineskip}

The authors declare no competing interests.

\bmhead{Supplementary information}

Supplementary Information contains complete principal-workflow prompts, retrieval-task definitions and complete initial and refined instructions for the FiQA case study (Supplementary Note 1), and key experimental settings (Supplementary Note 2; Supplementary Tables 1--2).

\clearpage
% ============================================================================
% Keep small table floats together without stretching them across separate pages.
\makeatletter
\setlength{\@fpsep}{2pt plus 2fil}
\makeatother
\section*{Extended Data}
\renewcommand{\tablename}{Extended Data Table}

% Extended Data tables follow their first citation in the main text and Methods.

\begin{table}[!htbp]
\caption{\textbf{Production-stage runtimes and finalised training-data sizes across 24 retrieval tasks.} Production timings are reported for one batch per listed task and query-source condition; when multiple batches were considered, the shorter-running batch was retained. Training records are valid entries in a finalised JSONL (JSON Lines) training-data file. Each batch has two equally sized training-file variants. Counts are reported per variant, without summing the variants or multiplying by training epochs. Timings include identifier extraction and attribute construction, query generation, candidate retrieval and final relevance annotation. Calibration, prebuilt indexing, failed attempts, waiting between resumed runs, final training-format conversion, and model training and evaluation are excluded. For resumed runs, elapsed times for the relevant completed stages are summed. For MS MARCO original queries, the reported runtime covers retrieval and annotation only. The measurements describe selected batches rather than task-average runtimes or total experimental costs. Data production used a system equipped with six NVIDIA GeForce RTX 5090 GPUs. Four hosted the generative model, and the other two were used only during embedding-model operations. Times are elapsed production-stage times, rather than GPU-hours or sums of per-GPU runtimes.}
\label{tab:ed-selected-production-runtime}
\centering
\small
\setlength{\tabcolsep}{6pt}
\begin{tabular}{lrr}
\toprule
Task & Training records & Production time (h) \\
\midrule
MS MARCO (original queries) & 9,546 & 4.444 \\
MS MARCO (SYNTRA queries) & 9,486 & 5.789 \\
LeCaRDv2 \cite{li2024lecardv2} & 8,278 & 63.786 \\
FiQA & 9,360 & 6.489 \\
SCIDOCS & 9,742 & 7.197 \\
SciRepEval Search & 9,302 & 7.040 \\
CQADupStack-Stats & 9,620 & 10.625 \\
Natural Questions & 3,886 & 2.429 \\
CQADupStack-Physics & 9,165 & 4.157 \\
NFCorpus & 2,897 & 2.250 \\
EvidenceBench & 9,817 & 7.223 \\
MedicalQA (1 query/document) & 1,919 & 2.498 \\
MedicalRetrieval & 9,754 & 3.561 \\
CodeTrans-DL (1 query/document) & 615 & 4.214 \\
CosQA & 5,140 & 3.761 \\
Text2SQL & 8,419 & 5.300 \\
BIRCO-relic & 4,909 & 4.614 \\
TheoremQA (questions) & 9,006 & 8.529 \\
TheoremQA (theorems) & 9,517 & 8.950 \\
CQADupStack-Mathematica & 9,337 & 8.245 \\
Touche2020 & 9,511 & 8.214 \\
ArguAna & 8,528 & 9.702 \\
DBpedia-entity & 9,537 & 6.662 \\
SpartQA & 1,062 & 5.381 \\
WinoGrande & 2,672 & 4.117 \\
\bottomrule
\end{tabular}
\end{table}

\begin{table}[!htbp]
\caption{\textbf{Agreement between SYNTRA-annotated relevance labels and human reference labels overall and by dataset.} Human reference labels were derived from seven annotators as described in Methods. Agreement is the fraction of exact matches, and $\kappa$ denotes Cohen's $\kappa$. Three-class evaluation distinguishes positives, hard negatives and easy negatives; binary evaluation merges the two negative classes. Macro-P, macro-R and macro-F1 denote class-averaged precision, recall and F1, respectively.}
\label{tab:rq2-agreement-summary}
\centering
\scriptsize
\setlength{\tabcolsep}{2.3pt}
\begin{tabular}{llrrrrrrr}
\toprule
Scope & Label space & $n$ & Matches & Agreement & $\kappa$ & Macro-P & Macro-R & Macro-F1 \\
\midrule
All samples & Three class & 150 & 103 & 0.687 & 0.510 & 0.713 & 0.749 & 0.699 \\
MS MARCO & Three class & 50 & 37 & 0.740 & 0.600 & 0.756 & 0.764 & 0.746 \\
Text2SQL & Three class & 50 & 32 & 0.640 & 0.444 & 0.742 & 0.824 & 0.704 \\
\shortstack[l]{TheoremQA\\(theorems)} & Three class & 50 & 34 & 0.680 & 0.484 & 0.670 & 0.725 & 0.675 \\
\midrule
All samples & Binary & 150 & 137 & 0.913 & 0.719 & 0.875 & 0.846 & 0.859 \\
MS MARCO & Binary & 50 & 44 & 0.880 & 0.705 & 0.893 & 0.829 & 0.851 \\
Text2SQL & Binary & 50 & 49 & 0.980 & 0.898 & 0.917 & 0.989 & 0.949 \\
\shortstack[l]{TheoremQA\\(theorems)} & Binary & 50 & 44 & 0.880 & 0.593 & 0.797 & 0.797 & 0.797 \\
\bottomrule
\end{tabular}
\end{table}

\begin{table}[!htbp]
\caption{\textbf{Confusion matrix and class-wise performance of SYNTRA-annotated relevance labels.} Human reference labels were derived from seven annotators as described in Methods. \textbf{a}, Three-class confusion matrix. Rows denote human reference categories and columns denote LLM-assigned categories. \textbf{b}, Class-wise performance. Human $n$ and LLM $n$ give the number of evaluation pairs assigned to each category. Among the 87 evaluation pairs labelled positive or hard negative by the LLM, 72 matched the human reference category, yielding a pooled precision of 82.8\%. The pooled precision combines the counts from both categories rather than averaging their precisions. These are the categories retained in the hard-negative-only training export.}
\label{tab:rq2-confusion}
\centering
\small
\setlength{\tabcolsep}{8pt}
\begin{tabular}{lrrr}
\multicolumn{4}{l}{\textbf{a. Three-class confusion matrix}} \\[2pt]
\toprule
Human reference & LLM positive & LLM hard negative & LLM easy negative \\
\midrule
Positive & 22 & 8 & 0 \\
Hard negative & 5 & 50 & 32 \\
Easy negative & 0 & 2 & 31 \\
\bottomrule
\end{tabular}
\par\medskip
\begin{tabular}{lrrrrr}
\multicolumn{6}{l}{\textbf{b. Class-wise performance}} \\[2pt]
\toprule
Class & Human $n$ & LLM $n$ & Precision & Recall & F1 \\
\midrule
Positive & 30 & 27 & 0.815 & 0.733 & 0.772 \\
Hard negative & 87 & 60 & 0.833 & 0.575 & 0.680 \\
Easy negative & 33 & 63 & 0.492 & 0.939 & 0.646 \\
\bottomrule
\end{tabular}
\end{table}

\begin{table}[!htbp]
\caption{\textbf{Data sizes and retrieval performance for the task-specific query-source comparisons.} \textbf{a}, Nonempty query entries submitted for relevance annotation, valid training records per positive-selection variant, and combined input rows supplied to training. Query entries include repeated query strings. Each valid record contains at least one positive and one hard negative. The two equally sized variants differ in positive-document selection as described in Methods and are concatenated without deduplication; combined input rows therefore equal twice the per-variant count. Input rows count supplied records before batching, rather than distinct queries or examples consumed over training epochs. \textbf{b}, nDCG@10 on a 0--100 scale, corresponding to Fig.~\ref{fig:results_analysis}d. The MS MARCO-trained model provides the shared baseline. Reference queries are from ReasonEmbed for the two TheoremQA tasks and official training pairs for CosQA and Synthetic Text2SQL. All task-specific query sources undergo SYNTRA relevance annotation and the shared training procedure.}
\label{tab:query-source-comparisons}
\centering
\small
\setlength{\tabcolsep}{4pt}
\begin{tabular}{llrrr}
\multicolumn{5}{l}{\textbf{a. Query and training-data sizes}} \\[2pt]
\toprule
Task & Query source & \shortstack[r]{Query\\entries} & \shortstack[r]{Records per\\variant} & \shortstack[r]{Combined\\input rows} \\
\midrule
TheoremQA (questions) & ReasonEmbed & 7,184 & 5,873 & 11,746 \\
 & SYNTRA & 9,852 & 9,006 & 18,012 \\
TheoremQA (theorems) & ReasonEmbed & 8,861 & 6,947 & 13,894 \\
 & SYNTRA & 9,880 & 9,517 & 19,034 \\
CosQA & Official & 9,020 & 6,599 & 13,198 \\
 & SYNTRA & 5,938 & 5,140 & 10,280 \\
Synthetic Text2SQL & Official & 10,000 & 8,245 & 16,490 \\
 & SYNTRA & 9,956 & 8,419 & 16,838 \\
\bottomrule
\end{tabular}
\par\medskip
\begin{tabular}{lrrr}
\multicolumn{4}{l}{\textbf{b. Retrieval performance}} \\[2pt]
\toprule
Task & \shortstack[r]{MS MARCO\\baseline} & \shortstack[r]{Reference\\queries} & \shortstack[r]{SYNTRA\\queries} \\
\midrule
TheoremQA (questions) & 16.90 & 25.53 & 32.868 \\
TheoremQA (theorems) & 11.61 & 23.07 & 22.199 \\
CosQA & 32.67 & 34.89 & 35.599 \\
Synthetic Text2SQL & 56.33 & 62.97 & 62.24 \\
\bottomrule
\end{tabular}
\end{table}

\begin{table}[!htbp]
\caption{\textbf{Paired BEIR and NanoBEIR results for the 18 models used in the
consistency analysis.} Values are suite-level mean nDCG@10 scores on a 0--100
scale. Ranks are computed separately within each model family; rank 1 denotes
the highest score.}
\label{tab:nanobeir-beir-models}
\centering
\small
\setlength{\tabcolsep}{4.5pt}
\begin{tabular}{llrrrr}
\toprule
Model family & Model & BEIR & NanoBEIR & \shortstack[r]{BEIR\\rank} & \shortstack[r]{NanoBEIR\\rank} \\
\midrule
\multirow{9}{*}{Encoder-only}
 & snowflake-arctic-embed-l  & 55.980 & 65.360 & 1 & 2 \\
 & snowflake-arctic-embed-m  & 54.900 & 64.420 & 2 & 3 \\
 & snowflake-arctic-embed-xs & 50.150 & 61.170 & 8 & 7 \\
 & bge-large-en-v1.5         & 54.340 & 65.910 & 3 & 1 \\
 & bge-base-en-v1.5          & 53.230 & 63.760 & 4 & 4 \\
 & bge-small-en-v1.5         & 51.660 & 62.650 & 5 & 5 \\
 & e5-large-v2               & 50.560 & 61.550 & 6 & 6 \\
 & e5-base-v2                & 50.300 & 61.010 & 7 & 8 \\
 & e5-small-v2               & 49.040 & 60.960 & 9 & 9 \\
\midrule
\multirow{9}{*}{Decoder-only}
 & NV-Embed-v2                    & 63.211 & 71.480 & 1 & 2 \\
 & e5-mistral-7b                  & 57.072 & 66.250 & 8 & 7 \\
 & gte-Qwen2-7B-instruct          & 58.860 & 71.390 & 5 & 3 \\
 & Qwen3-Embedding-8B             & 62.758 & 71.600 & 2 & 1 \\
 & Qwen3-Embedding-4B             & 61.578 & 69.910 & 3 & 5 \\
 & Qwen3-Embedding-0.6B           & 55.523 & 65.620 & 9 & 8 \\
 & stella\_en\_1.5B\_v5          & 59.311 & 70.920 & 4 & 4 \\
 & jina-embeddings-v5-text-small  & 57.379 & 67.190 & 7 & 6 \\
 & gte-Qwen2-1.5B-instruct        & 58.289 & 65.020 & 6 & 9 \\
\bottomrule
\end{tabular}
\end{table}

\begin{table}[htbp]
\caption{\textbf{Rank consistency between NanoBEIR and BEIR within embedding-model families.}
Spearman's $\rho$ was computed from paired suite-level mean nDCG@10 scores.
Ranks were assigned separately within each family, with rank 1 denoting the
highest score. $P$ values are two-sided.}
\label{tab:nanobeir-beir-summary}
\centering
\footnotesize
\setlength{\tabcolsep}{10.0pt}
\begin{tabular}{lrrrr}
\toprule
Model family & $n$ & Spearman's $\rho$ & $P$ value & Within one rank \\
\midrule
Encoder-only & 9 & 0.933 & $2.0\times10^{-4}$ & 8/9 \\
Decoder-only & 9 & 0.817 & $7.0\times10^{-3}$ & 6/9 \\
\bottomrule
\end{tabular}
\end{table}

\FloatBarrier
\end{document}

\begin{center}
{\Large\bfseries Supplementary Information}\par
\medskip
{\large Self-improving data synthesis can scale embedding-model adaptation across domains and tasks}\par
\medskip
Zucan Lyu, Rui Tang, Jianlyu Chen, Peisheng Jin and Defu Lian
\end{center}

This Supplementary Information presents the complete prompts for the principal synthesis workflow, retrieval-task definitions, complete initial and refined instructions for the FiQA case study, and experimental configurations. Shared prompt text is displayed once. Annotation prompts use the candidate-list format with document identifiers used in the formal experiments. Supplementary Note 2 summarizes key experimental settings in Supplementary Tables~\ref{tab:si-training-settings} and \ref{tab:si-strong-scaling-settings}.

\tableofcontents
\clearpage
\phantomsection\addcontentsline{toc}{section}{Supplementary Note 1: Prompts and task-specific instructions}
\section*{Supplementary Note 1: Prompts and task-specific instructions}
\Needspace{4\baselineskip}
\phantomsection\addcontentsline{toc}{subsection}{Source and input conventions}
\subsection*{Source and input conventions}
Prompts are user messages unless a system role is specified. Braced lower-case fields denote runtime inputs; upper-case fields denote shared display blocks that must be expanded before a prompt is sent. Each block contains the complete text of its branch, rather than an excerpt. The repository at \url{https://github.com/zucanlv/SYNTRA} contains the prompt builders and input formatters. The accompanying \path{docs/supplementary/} files provide fully expanded templates, optional branches, the complete selected instruction records and registry descriptors.

Prompt wording and branch-specific schemas are preserved. Line wrapping and blank-line spacing are typographic. The accompanying machine-readable files provide fully expanded templates; the source builders specify exact formatting. Decorative pictographs in the source are omitted, and arrows are written as ASCII. Python JSON expressions denote the serialized runtime value, using \path{ensure_ascii=False} and \path{indent=2}, rather than literal code passed to the model.

The principal flow infers task requirements, adapts identifier/attribute and query-generation guidance, synthesizes queries, and refines a relevance-annotation instruction. The refined annotation instruction is subsequently reused as the system message for production annotation. Generating several queries from one source document and annotating several candidate documents for one query are separate operations.

\textbf{Input fields.} \path{query_type}, \path{doc_type} and \path{relevance_def} describe the task. \path{doc} supplies a source document. \path{task_instruction} is the task-specific query instruction. \path{annotation_instruction} or \path{evaluation_instruction} is the relevance-annotation instruction. \path{feedback} supplies evaluation feedback; \path{samples_json} and \path{samples_brief} supply trial outputs. \path{refinement_guidance} and \path{current_guidance} hold the identifier/attribute addendum. \path{identifiers_brief}, \path{attributes}, \path{plans} and \path{already_selected} supply planning inputs. \path{num_plans} specifies the requested number of new plans; query generation produces one query for each supplied plan.

\textbf{Few-shot and demonstration inputs.} \path{target_examples_text} and \path{examples_text} format labelled target-task examples. \path{few_shot_section} contains query-only or query--positive examples according to the caller. Labelled examples include a hard-negative field only when available. \path{demo_blocks} pairs the MS MARCO and ArguAna reference examples with their task definitions; \path{anchor_block} contains their definitions alone. References equal to the target task are omitted. The task-definition caller uses up to five examples from each reference task in stored order. The complete fixed inputs and formatting functions are available in \path{Main_Pipeline/Few_Shot_Example.py}, \path{Few_Shot_Formatter.py} and \path{HighLevel_Generator.py}.

\textbf{Trial and candidate inputs.} Identifier/attribute validation serializes document identifiers, document text, identifiers and attributes; refinement omits document text. Query-instruction validation and refinement sample up to five and eight trial query--document pairs, respectively. Annotation candidates retain their order. Multi-document calls use zero-based local \path{doc_id} values, which are mapped back to original identifiers after parsing.
\Needspace{4\baselineskip}
\phantomsection\addcontentsline{toc}{subsection}{Task definition}
\subsection*{Task definition}
\Needspace{7\baselineskip}
\subsubsection*{Prompt 1: Task-definition generation}\label{si-prompt-1}
Infer the query type, document type and relevance criterion from examples. Replace \path{CONTRASTIVE_ANALYSIS} with one of the following complete lines, according to whether the target-task examples include hard negatives.\par
Source: \path{Main_Pipeline/HighLevel_Generator.py}.\par
\begin{Verbatim}
You are an expert Information Retrieval task analyst. Your job is to precisely characterise an IR task from its labeled examples.

## Two-Shot Demonstrations
Study these examples carefully — each shows how few-shot examples map to an abstract high-level task definition:

{demo_blocks}

---

## Your Task: Extract the High-Level Task Definition for the target corpus

### Few-Shot Examples to Analyse
{target_examples_text}

---

### Step-by-Step Analysis Instructions

Work through the following three steps in your reasoning. Be specific and cite evidence from the examples above.

**Step 1 — Query Analysis**
- What is the average length of the queries (short/medium/long)?
- What is the vocabulary style (natural language, keyword-style, formal, argumentative, domain-specific)?
- What is the user intent (fact-seeking, opinion-seeking, argument-seeking, research-oriented, ...)?
- What is the domain (general web, biomedical, legal, debate, ...)?
- Summarise into a precise 2-5 word phrase for `query_type`.

**Step 2 — Document Analysis**
- What type/structure do the Positive documents have (news articles, scientific abstracts, debate arguments, forum answers, ...)?
- What domain/register are they in?
- How long are they on average?
- Summarise into a precise 2-5 word phrase for `doc_type`.

{CONTRASTIVE_ANALYSIS}

**Step 3 — Relevance Definition Synthesis**
Combine the above analysis into ONE precise sentence following this exact template:
"Given a query (<query_type>) and a document (<doc_type>), the document is relevant if <specific criterion based on the contrastive analysis>."

The sentence must:
- Fill in <query_type> and <doc_type> from Steps 1 & 2.
- State the exact criterion (not just "it answers the query") derived from the contrastive analysis.
- Be specific enough that a language model reading only this sentence would produce queries and documents matching the style in the examples above.

---

### Output Format (MANDATORY — valid JSON only, no Markdown fences)
After your step-by-step reasoning, output a single JSON object:
{
  "query_type": "2-5 word phrase",
  "doc_type": "2-5 word phrase",
  "relevance_definition": "Given a query (...) and a document (...), the document is relevant if ..."
}

Output the JSON as the LAST thing in your response. Do not include any text after the closing brace.
\end{Verbatim}
\textbf{Target examples with hard negatives:}\par
\begin{Verbatim}
- **Contrastive Analysis**: Compare each Positive document against its Hard Negative. Hard Negatives are topically related but NOT relevant — identify the precise criterion that distinguishes a relevant document from a hard-but-wrong one. This is the single most important signal for writing a precise relevance_definition.
\end{Verbatim}
\textbf{Target examples containing positives only:}\par
\begin{Verbatim}
- **Contrastive Analysis**: No Hard Negative examples are provided. Focus on what specific aspect of the Positive documents makes them the ideal match for the queries.
\end{Verbatim}
\Needspace{7\baselineskip}
\subsubsection*{Prompt 2: Task-definition validation}\label{si-prompt-2}
Evaluate the inferred definition against the examples.\par
Source: \path{Main_Pipeline/HighLevel_Generator.py}.\par
\begin{Verbatim}
You are a senior quality evaluator for an information retrieval data synthesis pipeline.

## Your Task
Evaluate whether the high-level task definition below correctly and precisely characterises the IR task evidenced by the few-shot examples for the target corpus.

## Few-Shot Examples
{examples_text}

## Proposed High-Level Task Definition
- query_type: "{query_type}"
- doc_type: "{doc_type}"
- relevance_definition: "{relevance_def}"

## Evaluation Criteria (check each rigorously)

1. **Query Type Accuracy**: Does `query_type` precisely describe the style, intent, and domain of the queries in the examples? It should not be too generic (e.g., "query") nor too narrow.

2. **Document Type Accuracy**: Does `doc_type` precisely describe the Positive documents' type, structure, and domain? It should distinguish these documents from documents in other IR tasks.

3. **Relevance Definition Precision**: Does `relevance_definition` correctly explain why each Positive document is relevant to its query? Test it mentally: if you applied this definition as a filter, would it accept all Positive documents and reject all Hard Negatives?

4. **Task Distinctiveness**: Could this definition be confused with another well-known IR task? If so, the definition is too vague.

5. **Actionability**: Is the definition specific enough that an LLM reading only this sentence would know what kind of queries to generate and what kind of documents to retrieve?

## Step-by-step Evaluation
Go through each example one by one. For each:
- Does the query fit `query_type`?
- Does the Positive document fit `doc_type` and satisfy `relevance_definition`?
- (If Hard Negative present) Does the Hard Negative correctly fail `relevance_definition`?

Record any per-example issues, then summarise.

## Output Format (MANDATORY — valid JSON only, no Markdown)
{
  "passed": true | false,
  "overall_score": 1-4,
  "feedback": "Concise summary of the main issues, citing specific examples. Or confirm quality if none.",
  "query_type_issues": "Specific issues with query_type, or 'none'.",
  "doc_type_issues": "Specific issues with doc_type, or 'none'.",
  "relevance_def_issues": "Specific issues with relevance_definition (false accepts, false rejects, vagueness), or 'none'.",
  "suggestions": "Concrete, actionable rewrites or additions to fix the identified issues."
}

IMPORTANT: A "passed" verdict (overall_score >= 3) means the definition is production-ready. Any systematic mismatch with the examples must set passed=false.
\end{Verbatim}
\Needspace{7\baselineskip}
\subsubsection*{Prompt 3: Task-definition refinement}\label{si-prompt-3}
Revise the definition using evaluation feedback.\par
Source: \path{Main_Pipeline/HighLevel_Generator.py}.\par
\begin{Verbatim}
You are a prompt engineering expert specialising in information retrieval task characterisation.

## Context
We have a high-level task definition for the target corpus that was evaluated and found to have issues.

## Few-Shot Examples (ground truth)
{examples_text}

## Current (Flawed) Definition
- query_type: "{query_type}"
- doc_type: "{doc_type}"
- relevance_definition: "{relevance_def}"

## Evaluation Feedback
{feedback}

## Quality Anchors (reference definitions for other corpora)
{anchor_block}

## Your Task
Rewrite the high-level task definition for the target corpus to fix ALL issues identified in the feedback.

## Requirements
1. **Fix every issue**: Address each problem explicitly mentioned in the feedback.
2. **Stay grounded**: Every claim in the definition must be directly supported by at least one example.
3. **Maintain specificity**: query_type and doc_type must be 2-5 word phrases. relevance_definition must be one precise sentence.
4. **Follow the template**: "Given a query (<query_type>) and a document (<doc_type>), the document is relevant if <specific criterion>."

## Output Format (MANDATORY — valid JSON only, no Markdown fences)
{
  "query_type": "2-5 word phrase",
  "doc_type": "2-5 word phrase",
  "relevance_definition": "Given a query (...) and a document (...), the document is relevant if ..."
}
\end{Verbatim}
\Needspace{4\baselineskip}
\phantomsection\addcontentsline{toc}{subsection}{Identifier and attribute guidance}
\subsection*{Identifier and attribute guidance}
\Needspace{7\baselineskip}
\subsubsection*{Prompt 4: Identifier/attribute generation}\label{si-prompt-4}
Generate document-grounded identifiers and independent query-attribute dimensions. \path{GUIDANCE_BLOCK} is empty before refinement; subsequently insert the following block at the displayed position.\par
Source: \path{Main_Pipeline/IdAttr_Generator.py}.\par
\begin{Verbatim}
You are a professional data synthesis and NLP expert. Your task is to analyze the document and produce two outputs for downstream diverse and high-quality query generation:
1) Document Identifiers (information anchors)
2) Query Attributes (style/format/constraint dimensions with options)

## Task Context (use this to guide your analysis)
- This document is a: {doc_type}
- Queries that will target this document are: {query_type}
- Relevance criteria: {relevance_def}
- Tailor both identifiers and attribute dimensions so they are maximally
  useful for generating diverse queries of the above type.

## Core Definitions (must follow strictly)
- Identifier (Information Anchor): An irreplaceable core entity/relationship extracted directly from the document. It serves as the anchor that diverse queries should target. Do NOT include redundant/irrelevant details. Do NOT use external knowledge.
  - Entity-type: concrete objects in the document (e.g., people, places, events, records/results).
  - Relationship-type: unique relationships/logic between entities stated or implied by the document (e.g., relations between people, places, events, records/results).
- Query Attribute: a non-core feature dimension controlling how queries are asked. Attributes should NOT introduce new core facts; they should steer query diversity while keeping identifiers as the information anchors.
  - Example dimensions (for inspiration only; you may adopt, adapt, or omit): reasoning depth, answer form, topic specificity. These are common dimensions, but not all documents or query types need every one.
  - Brainstorm step: Besides considering the above examples, brainstorm 5-7 more independent dimensions by carefully analyzing THIS document (domain, topic, structure, content type, and what kinds of queries would realistically target it). Each dimension must be independent (no overlap with others), practical for query generation, and have 2-4 mutually exclusive options in mind.
  - Selection step (reason step-by-step): First, analyze the Task Context and the Input Document. Then strictly filter the full pool (example-inspired + brainstormed, typically 8-10 candidate dimensions) **using the Scientific Approach below**, and select exactly 3-6 dimensions that are most useful for diversifying queries over THIS document. Output only these 3-6 selected dimensions in the final JSON.

## Critical: Scientific Approach to Attribute Design
Before finalizing your Identifiers and Attributes, you MUST perform a "Task Alignment Analysis". Answer these questions:
1. **WHAT makes this task unique?**
   - Identify specific patterns, challenges, or domain characteristics of {query_type}. (Do not just repeat the task context; explain what it means in practice).
2. **WHY is each selected attribute suitable?**
   - Explain the MATCH between the attribute and the task characteristics. How does it help generate high-quality queries?
3. **MULTIPLE OPTIONS TEST (Crucial):**
   - Does the attribute genuinely have multiple valid, mutually exclusive options for THIS specific task?
   - CRITICAL RULE: If the task context restricts an attribute to only ONE valid option, you MUST NOT select it! (e.g., If it's a reasoning-intensive task, "reasoning depth" should always be "high", so "reasoning depth" is NOT a valid diversifying attribute).
This is not a template to fill - this is how experts think.
Blindly applying standard methods without understanding WHY is not acceptable.

## High-Quality Query Attributes Criteria
1. Strictly align with the characteristics of {query_type}.
2. Each selected dimension must have 2-4 mutually exclusive, actionable options applicable to THIS document.

{GUIDANCE_BLOCK}
## Input Document
doc: {doc}

## Output Format Requirements (MANDATORY, machine-parsable, no deviation)
- Output MUST be a single valid JSON object (RFC 8259).
- Output MUST contain ONLY the JSON object: no Markdown, no code fences, no commentary, no extra keys, no trailing text.
- Use double quotes for all strings and keys. No trailing commas.
- The JSON must start with a single left brace and end with a single right brace.
- Note: In this prompt template you may see doubled braces (e.g., `{{` / `}}`) due to Python string escaping; your OUTPUT JSON must use normal single braces.

## Hard constraints
- identifiers: 3-10 items; no duplicates/overlap; each must be a standalone anchor.
- attributes: 3-6 dimensions total. Produce them by (1) using the example dimensions only as inspiration, (2) brainstorming from THIS document, then (3) filtering via the Scientific Approach. Each dimension unique; each must have 2-4 options.
- options must be mutually exclusive, actionable, and applicable to THIS document's domain/topic for practical query generation.

## JSON schema (fixed keys; output MUST follow exactly)
{
  "identifiers": [
    {
      "type": "entity" | "relationship",
      "identifier": "string (copied/derived from document only; no external info)",
      "label": "string (3-8 words; short unique label)"
    }
  ],
  "attributes": [
    {
      "dimension": "string (clear independent dimension name)",
      "options": ["string", "string", "string", "string (optional)"]
    }
  ]
}
\end{Verbatim}
\begin{Verbatim}
## Refined Guidance (from prompt calibration — follow strictly)
{refinement_guidance.strip()}
\end{Verbatim}
\Needspace{7\baselineskip}
\subsubsection*{Prompt 5: Identifier/attribute validation}\label{si-prompt-5}
Evaluate trial outputs for task alignment, grounding and usable diversity.\par
Source: \path{Main_Pipeline/IdAttr_Generator.py}.\par
\begin{Verbatim}
You are a senior quality evaluator for an information retrieval data synthesis pipeline.

## Your Task
Evaluate whether the generated **Identifiers** and **Attributes** across the sample documents below are well-suited for producing diverse, high-quality **{query_type}** queries over **{doc_type}** documents.

## Step-by-step guideline:
1. **Grasp the task.** Understand the context and what good {query_type} queries look like for this task. What characterizes {query_type} queries here?
2. **Learn from examples.** Use the few-shot examples to infer the STYLE, DEPTH, and FOCUS of real {query_type} queries. The identifiers and attributes you evaluate should be able to yield {query_type} queries that match these in character.
3. **Evaluate the samples one by one.** There are multiple sample documents; evaluate **each sample's identifiers and attributes individually** and record issues as you go:
    i. **Compare to the task.** For each sample in turn, check whether its identifiers and attributes align with what good {query_type} queries need: diversity, grounding in the document, and actionable dimensions. Note any mismatch for that specific sample.
    ii. **Apply the criteria.** For each sample, evaluate its identifiers and attributes against the evaluation criteria (Identifier Quality, Attribute-Task Alignment, Multiple Options Test, Coverage, Actionability) and decide whether they are fit for the task. If there are issues, record them clearly for that sample (e.g., "Sample 2 (doc_id X): attribute dimension Y should not have multiple options in this task—it has a fixed characteristic and must be flagged.").
    Keep a clear record of per-sample issues so you can aggregate them in Step 4.
4. **Summarize.** Based on the per-sample issues you noted in Step 3, write a concise summary of the main issues—or confirm quality if none. This summary will go to the prompt engineer for prompt refinement.
5. **Format.** Respond strictly in the required JSON output format.

## Task Context
- Query Type: {query_type}
- Document Type: {doc_type}
- Relevance Definition: {relevance_def}

## Few-Shot Examples (gold-standard query-document pairs for this task)
{few_shot_section}

## Evaluation Criteria (check each rigorously)
1. **Identifier Quality**: Are identifiers diverse, non-redundant, and grounded in the document? Do they cover the key information anchors that {query_type} would realistically target?
2. **Attribute-Task Alignment**: Do the attribute dimensions genuinely help diversify {query_type}? Would each dimension produce meaningfully different queries when its options are varied?
3. **Multiple Options Test**: Does every attribute dimension have 2-4 mutually exclusive options that are ALL valid for this task? If the task context effectively restricts a dimension to a single option, it MUST be flagged.
4. **Coverage**: Across all samples, is there a healthy variety of attribute dimensions, or do the samples all use near-identical dimensions?
5. **Actionability**: Can a query generator actually use these identifiers + attributes to write concrete, natural {query_type} queries?

## Sample IdAttr Outputs (generated by the current prompt)
{samples_json}

## Scoring Rubric (assign exactly one integer score 0–3)
> **Important**: The parenthetical examples (e.g., ...) in each score level are purely illustrative of the *type* of issue at that severity. Do not treat them as a checklist — base your score on the holistic pattern across **all five evaluation criteria above**, not just the dimensions that happen to be mentioned in the examples.

- **Score 3 — Production-Ready**: Identifiers are diverse, specific, and fully grounded in each document's content. All attribute dimensions are genuinely task-aligned with 2–4 mutually exclusive options that are ALL valid for this query type (Multiple Options Test passes for every dimension). Coverage across samples shows healthy variety — samples do not all share near-identical dimension sets. A query generator can immediately produce diverse, natural {query_type} queries without any revision.
- **Score 2 — Acceptable**: Mostly sound, with only minor isolated issues — e.g., 1–2 identifiers that are slightly generic, or one attribute option that is marginally weak. Crucially, **no Multiple Options Test violations** (this is a hard disqualifier for Score 2+), and no systemic dimension repetition across samples. A query generator can still produce useful, diverse queries; only small prompt tweaks are needed.
- **Score 1 — Needs Revision**: Systematic problems undermine query diversity or groundedness. This includes: any Multiple Options Test violation (one or more attribute dimensions effectively restricted to a single valid option), identifiers that are consistently ungrounded or generic across samples (not anchored to document-specific content), or near-identical dimension sets repeating across samples with little variety. A query generator would produce repetitive, low-quality queries.
- **Score 0 — Unusable**: Fundamentally broken. Identifiers are absent, fabricated, or entirely off-topic relative to the documents. Attribute dimensions are structurally invalid, nonsensical, or all collapse to a single valid option. The output cannot be used for query generation at all.

## Output Format (MANDATORY — valid JSON only, no Markdown)
Output a single JSON object:
{
  "overall_score": 0-3,
  "passed": true if overall_score >= 2 else false,
  "feedback": "Concise summary of the main issues (or confirmation of quality). Be specific: cite which dimensions/identifiers are problematic and WHY. The summary will be sent to the prompt engineer to refine the prompt.",
  "identifier_issues": "Specific issues with identifiers, or 'none'.",
  "attribute_issues": "Specific issues with attribute dimensions/options, or 'none'.",
  "suggestions": "Concrete, actionable suggestions for improving the generation prompt. Focus on what guidance to ADD, not what to remove."
}

IMPORTANT: Be constructively critical. A "passed" verdict (overall_score >= 2) means the outputs are ready for production use. If you see ANY Multiple Options Test violation or systematic identifier/coverage issue, assign score <= 1 and set passed=false.
\end{Verbatim}
\Needspace{7\baselineskip}
\subsubsection*{Prompt 6: Identifier/attribute refinement}\label{si-prompt-6}
Produce a reusable task-level addendum. For an empty cache, \path{PREVIOUS_GUIDANCE} is the sentence below; otherwise it is \verb|{current_guidance.strip()}|.\par
Source: \path{Main_Pipeline/IdAttr_Generator.py}.\par
\begin{Verbatim}
You are a prompt engineering expert specializing in information retrieval data synthesis.

## Context
We have a prompt that instructs an LLM to generate **Identifiers** (information anchors) and **Attributes** (query diversity dimensions) for documents. The generated IdAttr will be used downstream to produce diverse **{query_type}** queries over **{doc_type}** documents.

- Relevance Definition: {relevance_def}

## Problem
The current prompt was evaluated on sample documents and received the following feedback. Note: if this feedback contains **specific examples** (e.g., "Identifier X for doc Y is wrong because ..."), you must **abstract** from these examples into **task-level** lessons.

### Evaluation Feedback
{feedback}

### Sample Outputs (from the current prompt)
{samples_brief}

## Previous Guidance (the last addendum that was tried)
{PREVIOUS_GUIDANCE}

## Your Task (two mental steps — output only the final guidance)

**Step 1 — Summarize and abstract (do this mentally, do not output it):**
- Read the feedback and identify **what went wrong** in the examples.
- **Generalize**: For each type of issue, extract the **underlying principle or pattern** that would help the generator avoid that class of error on **any** document, not just the ones in the sample. Ask: "What rule or criterion would prevent this kind of mistake in general?"

**Step 2 — Write Refined Guidance:**
Produce a **Refined Guidance** addendum that encodes those **task-level, query-agnostic** principles. This guidance will be inserted into the generation prompt and must apply to **all** future IdAttr generation, not to specific docs or queries.

## Requirements for the Refined Guidance
1. **Query-agnostic and task-level only**: Do NOT include doc-specific or query-specific fixes (e.g., do not reference particular doc_id, query text, or "for the query about X do Y"). Write only **general rules** that improve IdAttr generation for the whole task.
2. **Derived from feedback**: Every recurring issue type in the feedback should be reflected as a **general principle** in the guidance.
3. **Preserve what works**: If parts of the previous guidance are effective, keep them.
4. **Specific and actionable**: Say exactly WHAT to prefer/avoid and WHY for this task type.
5. **Concise**: 100-300 words. Dense, high-signal, imperative tone (e.g., "Prioritize ...", "Avoid ...", "For {query_type}, ensure ...").
6. **Only guidance about identifiers and attributes generation**: No meta-commentary, no JSON, no Markdown fences.

## Output Format
Output ONLY the refined guidance text (plain text, no JSON wrapping, no Markdown fences). This text will be inserted verbatim into the generation prompt.
\end{Verbatim}
\begin{Verbatim}
There was no previous guidance that was tried before.
\end{Verbatim}
\Needspace{4\baselineskip}
\phantomsection\addcontentsline{toc}{subsection}{Query-generation instructions}
\subsection*{Query-generation instructions}
\Needspace{7\baselineskip}
\subsubsection*{Prompt 7: Instruction construction}\label{si-prompt-7}
Construct a task-specific query-generation instruction from few-shot query--positive document pairs.\par
Source: \path{Main_Pipeline/QueryInstr_Generator.py}.\par
\begin{Verbatim}
You are an expert in information retrieval and synthetic query synthesis.

## Task Context
- Query Type: {query_type}
- Document Type: {doc_type}
- Relevance Definition: {relevance_def}

## Few-Shot Examples — (Query <-> Positive Document relevance pairs)
Each example shows a gold-standard {query_type} and a real {doc_type} that is relevant to it.
Use these pairs to infer how a real query author formulates the query-side expression that makes the document relevant.

{few_shot_section}

## Your Task — Two phases

**Phase 1 — Per-example analysis (think step-by-step for EACH example):**
For each (document, query) pair above, reason through:
1. Who is the real-world entity producing this type of query? What is their role, context, and goal?
2. What query-side purpose or expression would make the positive document relevant, and how is it formulated by the query author?
3. Which document details are authentic to include in this task's queries, and which details should be included, omitted, or transformed according to the query-document relationship demonstrated by the few-shot examples?
4. Does the query reflect the authentic voice, purpose, and framing of the query author, according to the query-document relationship demonstrated by the few-shot examples?

**Phase 2 — Cross-example generalisation (the actual output):**
After analysing all examples, output **exactly two sections** in the following order:

---

### Synthesizer Role

Write 1-2 sentences describing the role an LLM should *adopt* when synthesising {query_type} queries for this task.
The role must:
- Name who the producer is (e.g. "an everyday internet user", "a competitive debater", "a biomedical researcher")
- Describe what they are doing and what their goal is (e.g. "typing a query into a search engine to find an answer", "formulating a claim to argue in a debate round")
- Be grounded in the patterns observed across ALL few-shot examples — do NOT invent a role not supported by the data

---

### Query Extraction Process

Write a short, bulleted list of concrete steps the LLM should follow when reading a new {doc_type} and synthesising a {query_type} from it.
Each step must:
- Be stated as a general rule applicable to *any* {doc_type} in this corpus — do NOT reference specific entities, dates, or facts from the example documents
- Guide the LLM to infer a plausible query-side purpose or expression that would make the document relevant from the *producer's perspective* (as defined in Synthesizer Role)
- Specify which document-derived signals should shape the query, which details should be included/omitted/generalized, and how to frame the query according to the query-document relationship demonstrated by the few-shot examples

---

## Hard constraints for BOTH sections
- Every sentence must apply to any new {doc_type} encountered in this task, not just the examples above.
- Do NOT quote, paraphrase, or reference specific content from the example documents.
- The two sections together must be immediately usable as part of a system prompt without further editing.

## Output Format
Output ONLY the two sections above, in order, with their exact headings (`### Synthesizer Role` and `### Query Extraction Process`).
No JSON, no preamble, no per-example commentary, no additional text.
\end{Verbatim}
\Needspace{7\baselineskip}
\subsubsection*{Prompt 8: Style guidance}\label{si-prompt-8}
Extract the query style and register required by the task.\par
Source: \path{Main_Pipeline/QueryInstr_Generator.py}.\par
\begin{Verbatim}
### Role
You are a Senior Query Synthesist and Style Analyst. Your expertise lies in analyzing text patterns to create precise, actionable instructions for generating high-quality synthetic queries.

### Task Context:
- Target Output (Query Type): {query_type}
- Source Input (Document Type): {doc_type}
- Relevance Criteria: {relevance_def}

### Few-Shot Example Queries:
{few_shot_section}

### Instruction:
Based on the task context and the patterns observed in the few-shot examples, synthesize a "Task-Specific Style Guide". This guide will serve as the "gold standard" for an LLM to generate high-quality {query_type} queries for {doc_type} documents.

### Output Requirements:
You must actively analyze and distill features from both the Task Context and the Few-Shot Example Queries above. You mustobserve, abstract, and summarize what makes these queries fit the task, so the resulting guide can steer high-quality query generation. Address the following:

1. Linguistic DNA: From the examples, infer and state the tone (e.g. inquisitive, keyword-heavy), structural syntax, typical length, and any other recurring linguistic traits.
2. Relevance Heuristics: From the task context and examples, identify which information patterns in a {query_type} should be emphasised to satisfy the relevance criteria ({relevance_def}).
3. Negative Constraints: List "anti-patterns"—phrasing, styles, or excessive detail that would harm data purity and should be avoided.

### Output Format:
Produce a compact, bulleted instruction set suitable for embedding in a system prompt for synthetic query generation. Organise your output under the three subheadings above: Linguistic DNA, Relevance Heuristics, and Negative Constraints (with bullet points listed under each). 
Output only these subheadings and bullet points in plain text; no other text.
\end{Verbatim}
\Needspace{7\baselineskip}
\subsubsection*{Prompt 9: Instruction validation}\label{si-prompt-9}
Evaluate whether generated queries satisfy task requirements and match the example style.\par
Source: \path{Main_Pipeline/QueryInstr_Generator.py}.\par
\begin{Verbatim}
You are a senior quality evaluator for an information retrieval data synthesis pipeline.

## Your Task
Evaluate whether the **generated queries** below match the expected style, tone, and characteristics of real **{query_type}** queries for this task.

## Step-by-step guideline:
1. **Understand the task.** Be clear on what good {query_type} queries look like in this setting and what distinguishes them.
2. **Evaluate the (document, query) pairs one by one.** For each pair, evaluate the generated query against ALL five criteria below and record issues as you go.
    i. **Compare to the reference set.** Does the query match the golden queries in style, tone, length, and the level of abstraction/specificity?
    ii. **Apply the criteria.** Record any issue clearly per pair (e.g., "Pair 3: over-specifies document details, too long").
3. **Summarize.** Provide a concise summary of the main issues — or confirm quality is acceptable. If citing a problematic pair, use it only as evidence; the feedback itself must generalize the underlying task-level issue rather than preserve sample-specific entities, dates, facts, or wording. This summary will be used to refine a reusable query synthesis instruction.
4. **Output.** Respond strictly in the required JSON output format.

## Task Context
- Query Type: {query_type}
- Document Type: {doc_type}
- Relevance Definition: {relevance_def}

## Few-Shot Golden Queries (style + abstraction benchmark — compare generated queries to these)
{few_shot_queries_section}

## Current Query Synthesis Instruction (used to generate the queries below)
{query_instruction}

## Generated (Positive Document <-> Query) Pairs (sample)
{pairs_section}

## Evaluation Criteria (check each rigorously)
1. **Style Match**: Do the generated queries resemble the few-shot example queries in tone, length, vocabulary, and structure?
2. **Task Alignment**: Are the queries the kind of {query_type} that would realistically be issued against {doc_type} documents?
3. **Naturalness**: Do the queries read like something a real producer of this query type would write, or do they sound artificial/templated?
4. **User Authenticity**: Does each query feel like it was genuinely produced by the role described in the Synthesizer Role section of the instruction — i.e., someone acting in that role *without* having access to the source document? The key question is whether they are *authentic to the producer's perspective and intent in this task*. Penalise queries that merely describe or echo document content rather than reflecting genuine producer intent.
5. **Instruction Effectiveness**: Does the current instruction successfully guide the LLM on all four criteria above, including the Synthesizer Role and Query Extraction Process sections?

## Scoring Rubric (assign exactly one integer score 0–3)
> **Important**: Base your score on the holistic pattern across **all five criteria** above.

- **Score 3 — Production-Ready**: Generated queries closely match the golden queries in style, authenticity, naturalness, and the query-document relationship demonstrated by the examples. The Synthesizer Role and Query Extraction Process sections guide the LLM to produce queries from the genuine producer's perspective.
- **Score 2 — Acceptable**: Queries are broadly on-style and authentic, with only minor isolated deviations from the golden examples' producer perspective or query-document relationship. Only small targeted adjustments are needed.
- **Score 1 — Needs Revision**: Systematic issues across most queries — e.g., consistently using the wrong producer role, wrong level of specificity, wrong register, wrong relationship to the positive document, or otherwise missing the authentic producer perspective demonstrated in the golden examples. The Synthesizer Role or Query Extraction Process section is missing, vague, or not guiding correctly.
- **Score 0 — Unusable**: Queries are fundamentally off-task, incoherent, or incompatible with the task's expected query type and query-document relationship.

## Output Format (MANDATORY — valid JSON only, no Markdown)
Output a single JSON object:
{
  "overall_score": 0-3,
  "passed": true if overall_score >= 2 else false,
  "feedback": "Concise summary of the main issues. Be specific: cite example pairs that are problematic and explain WHY.",
  "style_issues": "Specific style mismatches vs few-shot examples, or 'none'.",
  "authenticity_issues": "Whether queries fail to reflect the genuine producer perspective (e.g. just echo document content, wrong role, wrong intent), or 'none'.",
  "instruction_issues": "What the query instruction gets wrong or misses, or 'none'.",
  "suggestions": "Concrete, actionable suggestions for improving the query instruction."
}

IMPORTANT: A "passed" verdict (overall_score >= 2) means the queries are production-ready. If there is ANY systematic authenticity or style mismatch across multiple pairs, assign score <= 1 and set passed=false.
\end{Verbatim}
\Needspace{7\baselineskip}
\subsubsection*{Prompt 10: Instruction refinement}\label{si-prompt-10}
Revise the reusable query-generation instruction using feedback.\par
Source: \path{Main_Pipeline/QueryInstr_Generator.py}.\par
\begin{Verbatim}
You are a prompt engineering expert specializing in information retrieval query synthesis.

## Context
We have a "Query Synthesis Instruction" that is embedded as a system prompt to steer an LLM toward producing **{query_type}** queries for **{doc_type}** documents.

- Relevance Definition: {relevance_def}

## Few-Shot Golden Queries (the gold standard — queries MUST match this style and abstraction level)
{few_shot_queries_section}

## Problem
The current instruction was evaluated and received the following feedback:

### Evaluation Feedback
{feedback}

### Sample (Positive Document <-> Generated Query) Pairs (produced by the current instruction)
{pairs_section}

### Current Query Synthesis Instruction
{query_instruction}

## Your Task
Rewrite the Query Synthesis Instruction to fix all issues identified in the feedback. The rewritten instruction must produce queries that closely match the few-shot examples in style, tone, length, and domain conventions.

## Requirements
1. **Preserve what works**: If parts of the current instruction are effective, keep them.
2. **Fix what's broken**: Address every issue in the feedback — including authenticity mismatches where generated queries fail to reflect the genuine producer perspective (e.g., they just echo document content, adopt the wrong register, or lack the intent of the role described).
3. **Be specific**: Instead of "use appropriate tone", specify exactly what tone (e.g., "**Tone:** Analytical, assertive, and occasionally contentious to reflect a strong stance.").
4. **Update the Synthesizer Role section if needed**: If the feedback indicates the LLM is not adopting the right producer identity, sharpen the role description to better convey who is writing the query and why.
5. **Update the Query Extraction Process section if needed**: If the feedback indicates queries are not reflecting the authentic producer perspective, revise the extraction steps to more explicitly guide the LLM to think from the producer's viewpoint.
6. **Format**: Output as a compact, bulleted query style guidelines set directly embeddable as a system prompt. No JSON wrapping. Keep the same style and structure as the current query style guidelines above (e.g. same level of detail, sectioning, and bullet conventions).

## Output Format
Output ONLY the rewritten instruction text (plain text with bullet points and section headings). No JSON, no Markdown fences, no commentary.
\end{Verbatim}
\Needspace{4\baselineskip}
\phantomsection\addcontentsline{toc}{subsection}{Query synthesis}
\subsection*{Query synthesis}
The following blocks are shared by blueprint planning and query realization. Expand each named block exactly at its displayed position. Planning selects information needs; realization writes one query per supplied blueprint.\par
\Needspace{4\baselineskip}
\textbf{QUERY\_INPUTS:}\par
\begin{Verbatim}
**You will be given:**
1) A {doc_type} (the ONLY source of truth)
2) A set of identifiers (information anchors extracted from the document)
3) A set of query attribute dimensions (each has multiple options)
\end{Verbatim}
\Needspace{4\baselineskip}
\textbf{TASK\_CONTEXT:}\par
\begin{Verbatim}
**Task Context:**
- Target Output (Query Type): {query_type}
- Source Input (Document Type): {doc_type}
- Relevance Criteria: {relevance_def}

**Few-Shot Examples:**
{few_shot_section}
\end{Verbatim}
\Needspace{4\baselineskip}
\textbf{STYLE\_GUIDANCE:}\par
\begin{Verbatim}
**Task-Specific Style Guidance:**
{task_instruction}
\end{Verbatim}
\Needspace{4\baselineskip}
\textbf{SOURCE\_DOCUMENT:}\par
\begin{Verbatim}
**Input document:**
{doc}
\end{Verbatim}
\Needspace{4\baselineskip}
\textbf{JSON\_ONLY:}\par
\begin{Verbatim}
- Output MUST be a single valid JSON object (RFC 8259).
- Output MUST contain ONLY the JSON object: no Markdown, no code fences, no commentary, no extra keys, no trailing text.
\end{Verbatim}
\Needspace{4\baselineskip}
\textbf{JSON\_SYNTAX:}\par
\begin{Verbatim}
- Use double quotes for all keys and strings. No trailing commas.
- JSON must start with '{' and end with '}'.
\end{Verbatim}
\Needspace{4\baselineskip}
\textbf{BLUEPRINT\_FIELDS:}\par
\begin{Verbatim}
      "used_identifier": ["identifier string"],
      "used_attributes": {
        "dimension name": "one option",
        "dimension name": "one option"
      }
    }
  ]
}
\end{Verbatim}
\Needspace{7\baselineskip}
\subsubsection*{Prompt 11: Blueprint planning}\label{si-prompt-11}
Use the shared blocks above. The branch-specific instructions and output schema below are complete.\par
Source: \path{Main_Pipeline/DiverseQuery_Generator.py}.\par
\begin{Verbatim}
You are planning diverse search query "blueprints" for information retrieval.
{QUERY_INPUTS}

{TASK_CONTEXT}

**Your task:** generate exactly {num_plans} diverse BLUEPRINTS. Each blueprint contains:
- exactly 1 chosen identifier (copy EXACTLY from the provided identifiers list)
- ONE attribute group: choose exactly ONE option for EACH attribute dimension

**Important:** In this stage, DO NOT write any search queries. Only output the selections (identifiers + attributes).

{STYLE_GUIDANCE}

Use this guidance to inform your blueprint planning decisions.

**Hard constraints:**
- Each blueprint must contain exactly 1 identifier — no more, no less.
- Do NOT invent identifiers or attribute options.
- Avoid redundancy: do not repeat the same identifier+attribute-group combination.

**Diversity planning (across these {num_plans} blueprints, and considering any already selected blueprints):**
- Maximize coverage of different identifiers.
- For each attribute dimension, try to cover as many different options as possible.

**Already selected blueprints (DO NOT repeat these combinations):**
{json.dumps(already_selected, ensure_ascii=False, indent=2)}

{SOURCE_DOCUMENT}

**Identifiers (choose from these; copy identifiers exactly):**
{json.dumps(identifiers_brief, ensure_ascii=False, indent=2)}

**Attribute dimensions (choose exactly one option per dimension per blueprint):**
{json.dumps(attributes, ensure_ascii=False, indent=2)}

**Output format requirements (MANDATORY):**
{JSON_ONLY}
- Do NOT include any "already_selected" plans in your output (do not repeat them); output ONLY NEW plans in the "plans" object.
{JSON_SYNTAX}

**JSON schema (fixed keys; output MUST follow exactly):**
{
  "plans": [
    {
{BLUEPRINT_FIELDS}
\end{Verbatim}
\Needspace{7\baselineskip}
\subsubsection*{Prompt 12: Query realization from blueprints}\label{si-prompt-12}
Use the shared blocks above. The branch-specific instructions and output schema below are complete.\par
Source: \path{Main_Pipeline/DiverseQuery_Generator.py}.\par
\begin{Verbatim}
You are a world-class synthetic query writer for information retrieval.

{QUERY_INPUTS}
4) A list of BLUEPRINTS (each blueprint already specifies exactly 1 identifier and exactly one option per attribute dimension)

{TASK_CONTEXT}

**Your task:** Generate a list of {len(plans)} distinct queries, strictly adhering to the provided blueprints.

{STYLE_GUIDANCE}

**Task Execution:**
1. Iterate through the provided Blueprints one by one.
2. Apply the Task-Specific Style Guidance to determine the appropriate linguistic register and domain-specific terminology (e.g., professional, academic, or casual).
3. For each blueprint, craft a high-quality query that implicitly but clearly reflects all assigned Attributes and targets the specified Identifiers.
4. Ensure the query is a perfect marriage of the blueprint's rigid requirements and the domain's stylistic nuances.

**NO reselection rule (STRICT):**
- You MUST NOT change the blueprint selections.
- For each query item in the output, you MUST include the blueprint's used_identifier and used_attributes EXACTLY as separate JSON fields (for reference). Do NOT embed them inside the "query" text.

**Attribute expressibility rule (STRICT):**
- The query text must clearly reflect the chosen options for each dimension (observable from wording).
- Even if a blueprint presents a logical challenge, you must provide the most plausible grounded query without altering the constraints.

{SOURCE_DOCUMENT}

**Identifiers reference:**
{json.dumps(identifiers_brief, ensure_ascii=False, indent=2)}

**Attribute dimensions reference:**
{json.dumps(attributes, ensure_ascii=False, indent=2)}

**Blueprints (write one query per blueprint, in the same order):**
{json.dumps(plans, ensure_ascii=False, indent=2)}

**Output format requirements (MANDATORY):**
{JSON_ONLY}
{JSON_SYNTAX}

**JSON schema (fixed keys; output MUST follow exactly):**
{
  "queries": [
    {
      "query": "string (one query matching the task-specified query type)",
{BLUEPRINT_FIELDS}
\end{Verbatim}
\Needspace{4\baselineskip}
\phantomsection\addcontentsline{toc}{subsection}{Relevance-annotation instructions}
\subsection*{Relevance-annotation instructions}
\Needspace{7\baselineskip}
\subsubsection*{Prompt 13: Annotation-instruction construction}\label{si-prompt-13}
Construct the task-specific annotation instruction for a query and a list of candidate documents carrying \path{doc_id} values.\par
Source: \path{Main_Pipeline/AnnoInstr_Generator.py}.\par
\begin{Verbatim}
You are an expert LLM annotator for information retrieval evaluation. Create a concise system instruction for another LLM annotator to evaluate documents and classify them into three categories: positive, hard negative, or easy negative.

**Critical thinking guideline:** The instruction should push the annotator to reason from the specific query-document relationship in this task: identify what would truly satisfy the query, what only looks relevant, and why the candidate crosses or fails that boundary. Avoid generic relevance advice.

The system instruction should specify following parts clearly:
**1. Task Context:**
- Query Type: {query_type}
- Document Type: {doc_type}
- Relevance Definition: {relevance_def}

**2. Scoring & Classification System:**
The instruction must define a task-specific 0-3 absolute scoring rubric grounded in the Query Type, Document Type, and Relevance Definition above. Guide the another annotator to reason about whether each document satisfies the task's stated relevance criteria, and to distinguish strong satisfaction, valid-but-less-complete positive cases, related-but-not-relevant cases, and clearly irrelevant cases.

Keep this label mapping fixed:
- **Score 3 -> positive**: Strongly satisfies the task-specific relevance definition.
- **Score 2 -> positive**: Satisfies the task-specific relevance definition well enough to be a valid positive, but is not enough to be a score 3 document.
- **Score 1 -> hard_negative**: The document looks plausible for the query in this task, but fails the task-specific relevance definition.
- **Score 0 -> easy_negative**: Clearly unrelated or outside the task-specific relevance criteria.

**3. Hard Negative Guidance:** The instruction should explain hard negatives with respect to the task’s actual relevance criteria: they may appear plausible for this task, but they fail the definition of relevance.

**4. Input Format:**
- One query string
- List of candidate documents with identifiers (doc_id)

**5. Annotation Protocol:**
1. Process sequentially: Annotate each document exactly once in the order provided. Ensure no doc_id is duplicated or omitted in the final output.
2. Independent evaluation: Evaluate each document in isolation based strictly on its relationship with the query. Do not allow the quality or content of other documents in the list to influence the current score (avoid relative ranking; use absolute scoring).

**6. Output Format (valid JSON only):**
{"query": "original query",
 "annotations": [{"doc_id": "id", "score": 0-3, "classification": "positive|hard negative|easy negative", "reasoning": "brief"}],
 "positive_doc_ids": [...], "hard_negative_doc_ids": [...], "easy_negative_doc_ids": [...]}

The instruction should be concise, emphasize hard negative identification, and require valid JSON output.
Output ONLY the system instruction text. **DO NOT** wrap the instruction in a JSON object!
\end{Verbatim}
\Needspace{7\baselineskip}
\subsubsection*{Prompt 14: Calibration-feedback generation}\label{si-prompt-14}
After calibration labels are compared with the reference labels, convert the confirmed mismatches into actionable feedback for instruction refinement. The label comparison supplies \path{compare_details}; \path{mismatches_json} serializes its mismatches with \verb|ensure_ascii=False| and \verb|indent=2|.\par
Source: \path{Main_Pipeline/AnnoCalibration_Generator.py}.\par
\begin{Verbatim}
You are a senior prompt evaluator for an information retrieval annotation pipeline.

## Your Task
The calibration labels are already confirmed, and pass/fail has already been decided by exact label comparison. Do not re-grade the calibration set and do not decide whether the run passed.

Your job is to turn the confirmed misclassifications below into constructive feedback for rewriting the annotation instruction. The feedback must explain what the current instruction is causing the annotator to misunderstand and how the next instruction should prevent the same mistakes.

## Step-by-step guideline
1. **Understand the task boundary.** Read the query type, document type, and relevance definition. Identify what a document must actually do to count as positive for this task.
2. **Analyze each mismatch.** For every mismatch, compare the expected label with the predicted label. Use the query, document text, expected reasoning, and predicted reasoning to infer the likely failure mode.
3. **Group repeated failure patterns.** Infer the recurring failure modes from the mismatches themselves instead of forcing them into predefined error categories.
4. **Diagnose instruction gaps.** Explain which missing or weak parts of the current annotation instruction likely allowed these errors.
5. **Write actionable refinement guidance.** Suggest concrete wording-level improvements for the next instruction. Focus on task-specific relevance criteria and hard-negative boundaries, not generic advice like "be more careful".

## Task Context
- Query Type: {query_type}
- Document Type: {doc_type}
- Relevance Definition: {relevance_def}

## Current Annotation Instruction
{annotation_instruction}

## Deterministic Calibration Result
- Total labels: {compare_details.get('total', 0)}
- Correct labels: {compare_details.get('correct', 0)}
- Mismatches: {len(mismatches)}

## Confirmed Misclassifications
Each item contains the query/document pair, the confirmed expected label, the annotator's predicted label, and any available expected/predicted reasoning.
{mismatches_json}

## Feedback Goal
Write structured feedback that guides the instruction refiner's thinking. The fields below are not separate checkboxes; together they should form a coherent diagnosis of the global failure pattern.

Do not merely enumerate individual errors. Use specific mismatches as evidence, but convert them into generalizable guidance that can prevent the same class of mistakes on future query-document pairs. Avoid generic advice such as "be more careful" or "think harder".

## Output Format (MANDATORY — valid JSON only, no Markdown)
Output a single JSON object:
{
  "feedback": "Concise global summary of the main failure pattern. Cite representative mismatches only as evidence for the broader diagnosis.",
  "positive_issues": "Global issues involving positive classifications, especially cases where plausibility caused false positives; use 'none' if not relevant.",
  "negative_issues": "Global issues involving hard/easy negative distinctions or missed hard-negative boundaries; use 'none' if not relevant.",
  "instruction_gaps": "The missing, weak, or ambiguous instruction guidance that likely caused the observed error pattern.",
  "suggestions": "Concrete task-specific changes the next instruction should make to prevent this class of errors."
}
\end{Verbatim}
\Needspace{7\baselineskip}
\subsubsection*{Prompt 15: Annotation-instruction refinement}\label{si-prompt-15}
Revise the instruction using calibration feedback while preserving the candidate-list annotation contract. \path{annotation_prompt_mode} specifies the contract with document identifiers.\par
Source: \path{Main_Pipeline/AnnoInstr_Generator.py}.\par
\begin{Verbatim}
You are a prompt engineering expert specializing in information retrieval evaluation.

## Context
We have an annotation instruction that guides an LLM to evaluate one or more candidate documents for a query and classify each document as positive, hard_negative, or easy_negative.

- Query Type: {query_type}
- Document Type: {doc_type}
- Relevance Definition: {relevance_def}
- Annotation Prompt Contract: {annotation_prompt_mode}

## Problem
The current instruction was evaluated and received the following feedback:

### Evaluation Feedback
{feedback}

### Current Instruction
{annotation_instruction}

## Your Task
Rewrite the annotation instruction to fix all issues identified in the feedback. The rewritten instruction must produce more accurate and consistent classifications for the multi-document annotation setting.

## Requirements
1. **Preserve the scoring system**: Keep the 0-3 score and three-class classification (positive/hard_negative/easy_negative).
2. **Preserve the multi-doc output format**: Keep the JSON schema with annotations, positive_doc_ids, hard_negative_doc_ids, and easy_negative_doc_ids, and require exact doc_id echoing for each candidate.
3. **Preserve what works**: If parts of the current instruction are effective, keep them.
4. **Fix what's broken**: Address every issue in the feedback, especially misclassification patterns and weak relevance-boundary guidance.
5. **Add task-specific guidance**: Include concrete criteria specific to {query_type}/{doc_type} that help distinguish true positives from hard negatives.
6. **Be precise about boundaries**: Clearly define where positive ends and hard_negative begins for this task.
7. **Preserve the instruction style**: Keep the same style and structure as the current annotation instruction above (e.g. same level of detail, sectioning, and bullet conventions).

## Output Format
Output ONLY the rewritten annotation instruction (plain text). No JSON wrapping, no Markdown fences. This will be used as a system prompt for the annotator LLM.
\end{Verbatim}
\Needspace{4\baselineskip}
\phantomsection\addcontentsline{toc}{subsection}{Production relevance annotation}
\subsection*{Production relevance annotation}
\subsubsection*{Prompt 16: Annotation system and user messages}\label{si-prompt-16}
Source: \path{Main_Pipeline/Batch_Candidate_Annotator.py}. The refined task-specific annotation instruction becomes the system message.\par
\textbf{System message:}\par
\begin{Verbatim}
{evaluation_instruction}
\end{Verbatim}
\textbf{User message:}\par
\begin{Verbatim}
Query: {query}

Candidate documents (each with doc_id and doc):

{candidate_document_blocks}
\end{Verbatim}
Each \path{candidate_document_blocks} entry is \verb|[doc_id: {doc_id}]|, followed by a newline, the document text and two newlines. Local identifiers are zero-based positions in the candidate chunk.\par
\textbf{Parser-retry user message:} the original conversation is retained and the following repair request is added.\par
\begin{Verbatim}
Your output was not a valid JSON or was missing required fields. Please output ONLY the strictly valid JSON object with annotations, positive_doc_ids, hard_negative_doc_ids, and easy_negative_doc_ids as specified in the system instructions.
\end{Verbatim}
\Needspace{4\baselineskip}
\phantomsection\addcontentsline{toc}{subsection}{Optional branches in the accompanying code}
\subsection*{Optional branches in the accompanying code}
The complete expanded templates in \path{docs/supplementary/shared_prompt_templates.json} also include the query-generation ablation without identifiers (\path{DiverseQuery_Generator.py}), the alternative annotation-instruction LLM validator (\path{AnnoInstr_Generator.py}), and document-type filtering with or without positive prototypes (\path{DocType_Filter.py}). These have distinct functions or output contracts and are retained as separate templates. Input-rendering code and fixed demonstrations are supplied in the source files identified above.
\Needspace{4\baselineskip}
\phantomsection\addcontentsline{toc}{subsection}{Task descriptors and registered examples}
\subsection*{Task descriptors and registered examples}
The registry spans 24 retrieval tasks and contains 121 examples: five per task, except six for DBpedia-entity. Each entry below gives its query type, document type and complete relevance definition. These definitions are reproduced from the released registry. Task-definition prompts infer and refine requirements from examples supplied at runtime. Example-source metadata and available identifiers accompany the code in \path{docs/supplementary/task_registry_descriptors.json}.
\begingroup\fontsize{10}{11.5}\selectfont\setlength{\parskip}{3pt}
\Needspace{5\baselineskip}
\textbf{MS MARCO}\par
\textit{Query type:} Web query. \textit{Document type:} Web passage.\par
\textit{Relevance definition:} Given a query (web query) and a document (web passage), the document is relevant to the query if the critical concepts or theories discussed in the document can provide references for domain experts to draft an answer to the query.\par\smallskip
\Needspace{5\baselineskip}
\textbf{LeCaRDv2}\par
\textit{Query type:} Chinese Criminal Case Description. \textit{Document type:} Chinese Criminal Judgment.\par
\textit{Relevance definition:} Given a query (a Chinese criminal case description) and a document (a candidate Chinese criminal judgment), the document is relevant if legal experts consider it legally comparable to the query, taking into account the criminal charge and similarities in critical facts, constitutive elements of the offense, sentencing circumstances, procedural controversies, and overall legal relevance.\par\smallskip
\Needspace{5\baselineskip}
\textbf{FiQA}\par
\textit{Query type:} Personal finance or investment question. \textit{Document type:} Financial answer passage.\par
\textit{Relevance definition:} Given a query (personal finance or investment question) and a document (financial answer passage), the document is relevant if it substantially addresses the specific financial scenario, decision, or concept raised by the query and provides information that would materially help answer the user's financial information need.\par\smallskip
\Needspace{5\baselineskip}
\textbf{SCIDOCS}\par
\textit{Query type:} Scientific paper title. \textit{Document type:} Scholarly article passage.\par
\textit{Relevance definition:} Given a query (scientific paper title) and a document (scholarly article passage), the document is relevant if it describes a scientific work or foundational source that the query paper cites or would reasonably cite, including closely related prior work, methods, datasets, benchmark systems, theoretical background, or seminal work that directly supports, motivates, contextualizes, or is compared with the query paper.\par\smallskip
\Needspace{5\baselineskip}
\textbf{SciRepEval Search}\par
\textit{Query type:} Scientific literature search query. \textit{Document type:} Scientific paper title, abstract, venue, and publication year.\par
\textit{Relevance definition:} Given a query (a scientific literature search query) and a document (a scientific paper represented by its title, abstract, venue, and publication year), the document is relevant if it directly satisfies the scientific information need expressed by the query, such as identifying a requested work or addressing the specified research topic, method, phenomenon, dataset, population, intervention, or outcome. Venue and year may help resolve publication-specific constraints, but do not by themselves make an otherwise topically unrelated paper relevant. A document that merely shares broad terminology while studying a different question or failing the query's specific constraints is insufficient.\par\smallskip
\Needspace{5\baselineskip}
\textbf{CQADupStack-Stats}\par
\textit{Query type:} Statistics Q\&A question. \textit{Document type:} Statistics Q\&A forum post.\par
\textit{Relevance definition:} Given a query (Statistics Q\&A question) and a document (Statistics Q\&A forum post), the document is relevant if it asks or addresses the same underlying statistical question or problem as the query. It should express an equivalent or near-duplicate information need, even if the wording, notation, dataset, or concrete example differs. Merely mentioning the same statistical method, analytical tool, or broad topic is not sufficient unless the document helps answer the query's specific concern.\par\smallskip
\Needspace{5\baselineskip}
\textbf{Natural Questions}\par
\textit{Query type:} Natural-language factoid question. \textit{Document type:} Wikipedia passage.\par
\textit{Relevance definition:} Given a query (natural-language factoid question) and a document (Wikipedia passage), the document is relevant if it contains factual evidence that directly answers the question or provides the specific information needed to derive the answer. Merely discussing the same entity or broad topic without resolving the question is insufficient.\par\smallskip
\Needspace{5\baselineskip}
\textbf{CQADupStack-Physics}\par
\textit{Query type:} Physics Q\&A question. \textit{Document type:} Physics Q\&A forum post.\par
\textit{Relevance definition:} Given a query (Physics Q\&A question) and a document (Physics Q\&A forum post), the document is relevant if it asks or addresses the same underlying physics question or problem as the query. It should express an equivalent or near-duplicate information need, even if the wording, notation, physical system, or concrete example differs. Merely mentioning the same physical concept, theory, or formula is not sufficient unless the document helps answer the query's specific concern.\par\smallskip
\Needspace{5\baselineskip}
\textbf{NFCorpus}\par
\textit{Query type:} Non-technical health or nutrition query. \textit{Document type:} Scientific medical article or biomedical research abstract.\par
\textit{Relevance definition:} Given a query (non-technical health or nutrition query) and a document (scientific medical article or biomedical research abstract), the document is relevant if it provides scientific evidence, clinical or epidemiological findings, experimental results, or biomedical explanations that directly address or substantially inform the health, nutrition, disease, or dietary intervention topic raised by the query.\par\smallskip
\Needspace{5\baselineskip}
\textbf{EvidenceBench}\par
\textit{Query type:} Biomedical scientific hypothesis. \textit{Document type:} Sentence from a biomedical research paper.\par
\textit{Relevance definition:} Given a query (a biomedical scientific hypothesis) and a document (a sentence from a biomedical research paper), the document is relevant if it reports a finding, observation, analysis, or conclusion that directly supports, contradicts, qualifies, or otherwise provides evidence about the relationship or outcome asserted by the hypothesis. Merely mentioning the same biomedical entities or broad topic is insufficient.\par\smallskip
\Needspace{5\baselineskip}
\textbf{MedicalQA}\par
\textit{Query type:} English-language medical question. \textit{Document type:} English-language medical answer passage.\par
\textit{Relevance definition:} Given a query (an English-language medical question) and a document (an English-language medical answer passage), the document is relevant if it directly provides the specific medical information needed to answer the question, such as information about a disease, condition, symptom, drug, medical test, procedure, or treatment. Merely mentioning the same medical topic or entity without addressing the specific aspect asked about in the query is insufficient.\par\smallskip
\Needspace{5\baselineskip}
\textbf{MedicalRetrieval}\par
\textit{Query type:} Chinese-language medical or health question. \textit{Document type:} Chinese-language medical answer passage.\par
\textit{Relevance definition:} Given a query (Chinese-language medical or health question) and a document (Chinese-language medical answer passage), the document is relevant if it provides a direct and medically pertinent answer to the question or otherwise helps resolve the health concern expressed in the query.\par\smallskip
\Needspace{5\baselineskip}
\textbf{CodeTrans-DL}\par
\textit{Query type:} TensorFlow Code Snippet. \textit{Document type:} PaddlePaddle Code Snippet.\par
\textit{Relevance definition:} Given a query (TensorFlow Code Snippet) and a document (PaddlePaddle Code Snippet), the document is relevant if it implements the same deep learning functionality as the query using PaddlePaddle APIs and conventions, making the two snippets functionally equivalent cross-framework counterparts.\par\smallskip
\Needspace{5\baselineskip}
\textbf{CosQA}\par
\textit{Query type:} Natural language web query for code search. \textit{Document type:} Python function implementation with optional documentation/comments.\par
\textit{Relevance definition:} Given a query (Natural language web query for code search) and a document (Python function implementation with optional documentation/comments), the document is relevant if the code implementation and its internal documentation (docstrings) provide a direct, functional solution to the specific programming task or logic described in the query.\par\smallskip
\Needspace{5\baselineskip}
\textbf{Text2SQL}\par
\textit{Query type:} Natural Language Database Query. \textit{Document type:} SQL Query.\par
\textit{Relevance definition:} Given a query (Natural Language Database Query) and a document (SQL Query), the document is relevant if it accurately and completely implements the database operation described in the query.\par\smallskip
\Needspace{5\baselineskip}
\textbf{BIRCO-relic}\par
\textit{Query type:} Literary analysis excerpt with a masked quotation. \textit{Document type:} Candidate passage from a literary work.\par
\textit{Relevance definition:} Given a query (an excerpt of literary analysis containing a quotation replaced by '[masked sentence(s)]') and a document (a candidate passage from the analyzed literary work), the document is relevant if it can be naturally inserted at the masked position and directly supports at least one of the literary claims made in the surrounding context. Sharing the same work, characters, themes, or vocabulary without supporting the surrounding interpretation, or merely repeating content already present in the query, is insufficient.\par\smallskip
\Needspace{5\baselineskip}
\textbf{TheoremQA questions}\par
\textit{Query type:} Theorem-based STEM word problem. \textit{Document type:} Solved STEM problem.\par
\textit{Relevance definition:} Given a query (theorem-based STEM word problem) and a document (solved STEM problem), the document is relevant if it uses the same or a closely related theorem, principle, formula, or reasoning pattern required to solve the query, providing a worked example that can guide the solution.\par\smallskip
\Needspace{5\baselineskip}
\textbf{TheoremQA theorems}\par
\textit{Query type:} Mathematical word problem. \textit{Document type:} Mathematical theorem, definition, or proof.\par
\textit{Relevance definition:} Given a query (mathematical word problem) and a document (mathematical theorem, definition, or proof), the document is relevant if it states, defines, or proves a theorem or mathematical principle that is used to derive the solution to the query.\par\smallskip
\Needspace{5\baselineskip}
\textbf{CQADupStack-Mathematica}\par
\textit{Query type:} Mathematica Q\&A question. \textit{Document type:} Mathematica Q\&A forum post.\par
\textit{Relevance definition:} Given a query (Mathematica Q\&A question) and a document (Mathematica Q\&A forum post), the document is relevant if it asks or addresses the same underlying Mathematica or Wolfram Language question or problem as the query. It should express an equivalent or near-duplicate information need even if the wording, code, data, or concrete example differs. Merely mentioning the same function, symbol, or broad topic is insufficient unless the document helps resolve the query's specific concern.\par\smallskip
\Needspace{5\baselineskip}
\textbf{Touche2020}\par
\textit{Query type:} Controversial Argumentative Question. \textit{Document type:} Argumentative passage with reasoning.\par
\textit{Relevance definition:} Given a query (Controversial Argumentative Question) and a document (Argumentative passage with reasoning), the document is relevant if it provides a comprehensive, logically structured argument or rebuttal that takes a clear stance on the topic, supporting its position with specific contentions, evidence, or reasoning.\par\smallskip
\Needspace{5\baselineskip}
\textbf{ArguAna}\par
\textit{Query type:} Argumentative Claim. \textit{Document type:} Counter-argument Passage.\par
\textit{Relevance definition:} Given a query (argumentative claim) and a document (counter-argument passage), the document is relevant if it provides a valid, logically structured counter-argument that directly challenges, refutes, or offers a conflicting perspective to the central premise of the query.\par\smallskip
\Needspace{5\baselineskip}
\textbf{DBpedia-entity}\par
\textit{Query type:} Entity-centric search query. \textit{Document type:} Encyclopedic entity description.\par
\textit{Relevance definition:} Given a query (entity-centric search query) and a document (encyclopedic entity description), the document is relevant if it provides a factual summary of a specific entity, instance, or concept that directly satisfies the categorical constraints and attributes specified in the query.\par\smallskip
\Needspace{5\baselineskip}
\textbf{SpartQA}\par
\textit{Query type:} Textual spatial scene followed by a multiple-choice question. \textit{Document type:} Candidate spatial answer phrase.\par
\textit{Relevance definition:} Given a query (a textual spatial scene followed by a multiple-choice question) and a document (a candidate spatial answer phrase), the document is relevant if it denotes an answer supported by the scene's stated spatial relations and valid inferences across objects and containers. If both choices satisfy the queried relation, each choice and 'both of them' are relevant; if neither does, 'none of them' is relevant. Mere object mention or lexical overlap is insufficient.\par\smallskip
\Needspace{5\baselineskip}
\textbf{WinoGrande}\par
\textit{Query type:} Commonsense fill-in-the-blank sentence. \textit{Document type:} Candidate answer phrase.\par
\textit{Relevance definition:} Given a query (a sentence containing one blank marked by an underscore) and a document (a candidate answer phrase), the document is relevant if inserting it into the blank yields the intended completion under the causal, physical, social, or temporal commonsense constraints of the full sentence. Grammatical compatibility or lexical association alone is insufficient.\par\smallskip
\endgroup
\Needspace{4\baselineskip}
\phantomsection\addcontentsline{toc}{subsection}{FiQA case study: complete initial and refined instructions}
\subsection*{FiQA case study: complete initial and refined instructions}
This section provides the complete instruction pairs and original examples underlying the FiQA case study in main-text Fig. 5. Query-instruction refinement directed generation towards a user's financial information need, while annotation-instruction refinement clarified that correcting a query's premise can be relevant. The instruction texts below are reproduced in full from the saved records.
\subsubsection*{Query-generation example and instructions}\label{si-fiqa-query}
The initial and refined instructions correspond to the first and third recorded attempts. The source document and queries below are verbatim production examples generated in separate runs using these instructions. Attribute plans differed between the runs.\par
\textbf{Source document (doc\_id 310844):}\par
\begin{Verbatim}
There's no catch. Banks need to acquire customers just like any other business. One common way to acquire new customers is by advertising on the radio, TV, print, etc. Another common way to acquire new customers is by offering incentives like the one you linked to. Basically, PNC is confident that they will make more than $100 in profit over the entire lifetime of a customer. This is a very reasonable assumption, considering that:
\end{Verbatim}
\textbf{Query generated with the initial instruction:}\par
\begin{Verbatim}
Why is PNC offering $100 to new customers if they're confident they'll make more than $100 in profit over the lifetime of that customer?
\end{Verbatim}
\textbf{Query generated with the refined instruction:}\par
\begin{Verbatim}
why is pnc giving $100 just for opening an account
\end{Verbatim}
\textbf{Initial-instruction plan:}\par
\begin{Verbatim}
{"used_identifier": ["PNC"], "used_attributes": {"query focus": "bank incentive strategy", "user perspective": "individual consumer", "answer form": "explanation of rationale"}}
\end{Verbatim}
\textbf{Refined-instruction plan:}\par
\begin{Verbatim}
{"used_identifier": ["PNC"], "used_attributes": {"query focus": "bank incentive strategy", "financial decision type": "personal banking choice", "reasoning depth": "basic factual inquiry"}}
\end{Verbatim}
\Needspace{5\baselineskip}
\textbf{Complete initial query-generation instruction (attempt 1):}\par
\begin{Verbatim}
### Synthesizer Role

The synthesizer should adopt the role of an individual seeking practical financial clarity, acting as a retail investor, taxpayer, or household budgeter. Their goal is to type concise, direct questions into a search engine or community forum to resolve specific points of confusion, compare financial strategies, or understand the practical implications of a financial rule.

### Query Extraction Process

*   **Identify the Core Financial Dilemma:** Pinpoint the specific problem, trade-off, or conceptual confusion the passage addresses, such as a choice between two investment vehicles, a misunderstanding of a fee, or the tax consequences of an action.
*   **Determine the User’s Motivation:** Infer whether the user is looking for a "how-to" guide (reducing expenses), a "what’s the difference" comparison (trading vs. flipping), or a "can I do this" permission-based inquiry (using short-sale proceeds).
*   **Adopt a Direct, Natural Voice:** Frame the query using everyday financial vocabulary; avoid overly formal or academic language in favor of the terminology a non-expert would likely use.
*   **Abstract Specifics into Scenarios:** If the document provides a detailed numerical example or specific case study, generalize it into a broader situational query while retaining the essential financial variables.
*   **Focus on the Underpinning Intent:** Ensure the query strikes at the heart of the document’s unique insight rather than just matching keywords; if the document debunks a myth or offers a psychological tip, frame the query to elicit that specific guidance.
*   **Maintain Concision:** Limit the query to a single sentence or a brief phrase that captures the primary information need, omitting any extraneous details from the passage that the user would not yet know.

### Linguistic DNA
*   **Varied Intent Syntax:** Queries should range from direct "How-to" questions and "What is" definitions to complex situational scenarios and mechanical "Is it possible" inquiries.
*   **Pragmatic Tone:** The tone must be inquisitive and utilitarian, reflecting an individual seeking actionable financial advice or clarification on technical market rules.
*   **Variable Length:** Synthesize queries across a spectrum of brevity, from short keyword-based phrases (4-6 words) to detailed multi-part scenarios (15-25 words).
*   **Domain-Specific Vernacular:** Use common financial terminology naturally (e.g., "401k," "shorting," "tax implications," "long position," "initial payment").
*   **Informal Polish:** Maintain a style that mimics real-world user behavior—grammatically functional but often lacking perfect capitalization or formal academic structure.

### Relevance Heuristics
*   **Scenario-Action Coupling:** Focus on queries that pair a specific financial situation (e.g., being in your early 20s) with a specific decision point (e.g., 401k investment choices).
*   **Mechanism Inquiries:** Emphasize the "how" of financial movement, such as the liquidity of funds between trades or the timing of down payments.
*   **Comparative Analysis:** Generate queries that require the document to distinguish between two related but distinct concepts (e.g., day trading vs. flipping).
*   **Outcome Orientation:** Ensure the query seeks a material benefit, such as spending less, reducing tax liability, or accelerating a home purchase timeline.
*   **Constraint-Based Questions:** Frame queries around limitations or permissions within financial regulations and account types.

### Negative Constraints
*   **No Generic Financial Literacy:** Avoid overly broad "What is money?" or "What is the stock market?" queries that lack a specific user intent or scenario.
*   **No Hyper-Dense Jargon:** Avoid queries that read like textbook headings; they should feel like a human asking for help, not a search engine index.
*   **No Opinion-Seeking:** Avoid subjective prompts like "What is the best stock?" or "Do you like Vanguard?"; focus on factual, procedural, or strategic information needs.
*   **No Excessive Preamble:** Eliminate "filler" phrases like "I was wondering if you could tell me..." Start directly with the core question or scenario.
*   **No Unrelated Personal Anecdotes:** Keep scenarios concise; do not include irrelevant biographical details that do not impact the financial logic of the question.
\end{Verbatim}
\Needspace{5\baselineskip}
\textbf{Complete refined query-generation instruction (attempt 3):}\par
\begin{Verbatim}
### Synthesizer Role
You are a person with a specific money problem searching for an immediate answer on a community forum (like Reddit) or a mobile search bar. You are the "Uninformed Real-World Actor"—a retail investor, a stressed spouse, or a first-time homebuyer. You do not have the document's expert knowledge. You have a "Minimum Viable Inquiry" mindset: you type only the necessary words to find a solution. You are pragmatic, blunt, and often use shorthand. You never act like a student or a researcher; you act like someone whose bank balance is at stake.

### Query Synthesis Methodology
*   **The "Pre-Knowledge" Shift:** Before writing, identify the technical jargon in the document and consciously forget it. If the document says "Capital Gains," you think "profit tax." If it says "Amortization," you think "monthly payment."
*   **The Problem-Source Pivot:** Do not summarize what the document says. Instead, imagine the messy, real-life scenario that forced someone to look for this information. (e.g., Document: "KYC Regulations." User: "why does the bank need my ID?")
*   **Translate Jargon to Street-Talk:** Use the most common, non-technical equivalent of words. Replace "Leave Encashment" with "getting paid for unused vacation"; replace "Identity Verification" with "bank asking for my ID."
*   **The Financial Forced-Angle:** If a document is non-financial (e.g., about psychology or hobbies), you MUST extract or invent a financial motive. Focus exclusively on cost, price, or saving money. (e.g., If the document is about free vs paid content, the query is "how to spend less on subscriptions" or "why pay for it when its free.")
*   **Identify the Core Conflict:** Search for a "this vs. that" or a "can I do [Action X]" logic. Use the document as a "Yes" or "No" map for a specific user action.

### Linguistic DNA & Style
*   **Short & Blunt:** Keep queries between 5-10 words. No complex clauses.
*   **Search Bar Syntax:** Use sentence-case or all lowercase. Avoid perfect punctuation.
*   **Contractions & Shorthand:** Use "401k," "IRS," "crypto," "stock," "tax," "house." Use "whats" instead of "what is."
*   **Zero Preamble:** Never use "Can you," "I need to know," or "I am looking for." Start directly with the noun or the verb. (e.g., "trading vs flipping taxes" NOT "What is the difference between...")
*   **Register:** Utilitarian, informal, and impatient.

### Failure Mode Contrast (Instructional Examples)
*   **Too Technical (FAIL):** "What are the tax implications of leave encashment at retirement?"
*   **Authentic (PASS):** "do i pay tax on unused vacation payout when I quit"
*   **Too Clinical (FAIL):** "Is there a financial benefit to utilizing free internet resources?"
*   **Authentic (PASS):** "how can i spend less on this" or "is it worth paying for"
*   **Echoing (FAIL):** "Know your customer identification requirements for bank accounts."
*   **Authentic (PASS):** "why the bank needs my id just to open an account"

### Negative Constraints
*   **FORBID Specific Nouns:** Do not use the specific technical titles or headers found in the document. If the title is "Amortization Schedule," you are forbidden from using "Amortization" or "Schedule."
*   **NO Self-Labeling:** Never say "As a 20 year old..." or "I am a beginner." Incorporate context naturally: "best 401k for early 20s" or "how to start investing."
*   **NO Summarization:** Do not try to encompass the whole document. Target one specific, actionable question that the document answers.
*   **NO Academic Words:** Ban words like: "implications," "considerations," "factors," "pertaining," "regarding," "significant," or "process."

### Final Quality Filter
Does this query look like it was typed into a phone with one hand while standing in a checkout line?
*   Gold Standard Example: "Whats the difference between day trading and flipping and their tax implications?" (Note: "Tax implications" is the only technical term allowed because it is a common search-engine shorthand).
*   Gold Standard Example: "Buying a House and Taking Part of 20% to move initial payment date forward"
\end{Verbatim}
\subsubsection*{Relevance-annotation example and instructions}\label{si-fiqa-annotation}
The following known-positive few-shot pair is the original example summarized in Fig. 5b. The initial instruction corresponds to the first recorded attempt; the refined instruction corresponds to the second. The recorded comparison changes from \path{hard_negative} to \path{positive}, matching the reference label after refinement.\par
\Needspace{4\baselineskip}
\textbf{Original query:}\par
\begin{Verbatim}
Buying a House and Taking Part of 20% to move initial payment date forward
\end{Verbatim}
\textbf{Original positive document:}\par
\begin{Verbatim}
In the prior PMI discussions here, it's been stated that the bank is not obligated to remove PMI until the mortgage's natural amortization puts the debt at 78% LTV. So, paying in advance like this will not automatically remove the PMI.  Nor will a lump sum payment be certain to move the next payment ahead a year. If it's entered as a principal prepayment, the next month's payment is still due. In the world of coupon books, if you sent in a year's payments, you'd not benefit from the interest saved, in one year you'd owe what the amortization table tells you.  There's no free lunch when it comes to mortgages or finance in general. This is why we usually caution that one should not be cash poor the day after buying a house. Best to save 30%, put down 20%, and have a cushion after the closing.
\end{Verbatim}
\Needspace{5\baselineskip}
\textbf{Complete initial relevance-annotation instruction (attempt 1):}\par
\begin{Verbatim}
You are a financial relevance annotator. For each query-document pair, judge whether the document substantially addresses the specific financial scenario, decision, or concept raised by the query and provides information that would materially help answer the user’s financial information need.

**Scoring Rubric (Absolute 0–3)**
- **Score 3 – Positive**: Directly and fully addresses the query’s precise financial scenario, decision, or concept. Contains the critical details, conditions, or actionable guidance the user actually needs. No essential dimension is omitted.
- **Score 2 – Positive**: Adequately addresses the core need but is less complete, less specific, or less authoritative than a Score 3. It still provides material help—just missing some nuance, depth, or coverage of a secondary aspect. Valid positive, but not the strongest answer.
- **Score 1 – Hard Negative**: Plausible at a glance (shares keywords, context, or the broad topic) but fails the relevance definition. It does **not** materially help answer the query because it either misses the specific financial scenario/decision, addresses a different angle, omits the decisive constraint, or provides information that is tangential, outdated, or inapplicable to the user’s situation. Only score 1 when a reasonable reader could initially think it might be relevant, but careful analysis shows it isn’t.
- **Score 0 – Easy Negative**: Clearly unrelated, off-topic, or outside the task’s relevance criteria. No plausible connection to the query.

**Hard Negative Reasoning (Score 1)**  
To identify hard negatives, ask: “What is the exact financial information need?” and “What makes this document look relevant, yet fail?”  
Common hard-negative patterns in this task:
- **Topic overlap, wrong decision context**: The document discusses the same general instrument or concept (e.g., “Roth IRA”), but the query asks about a specific strategy, tax treatment, or eligibility condition that the document never addresses (e.g., query: “tax implications of a backdoor Roth IRA” -> document explains the basics of a Roth IRA without mentioning backdoor conversions or their tax nuances).
- **Partial match missing the operative constraint**: The query includes a critical qualifier (e.g., “for self-employed,” “in retirement,” “for crypto”) that materially changes the answer. The document provides generic information without that qualifier, rendering the response inapplicable or misleading.
- **Wrong temporal or jurisdictional scope**: The document is about an outdated rule, a different country’s regulation, or a repealed provision when the query implies current, actionable guidance.
- **Looks helpful but solves the wrong problem**: The document explains how to perform an action, but the query asks whether the action is advisable in a specific context, or asks for comparison/alternatives the document never covers.

In your reasoning, explicitly state which part of the query’s need the document fails to meet, and why a surface-level similarity is misleading.

**Input Format**  
You will receive:  
- A query string (personal finance or investment question)  
- A list of candidate documents, each with a unique `doc_id`.

**Annotation Protocol**  
1. Process documents in the given order. Annotate every document exactly once; no duplication or omission.  
2. Evaluate each document independently. Do not compare documents; assign an absolute score based solely on the query-document relationship.  
3. Apply the rubric strictly. If in doubt between Score 1 and Score 2, re-read the query’s specific scenario/decision and check whether the document’s information materially advances an answer.

**Output Format**  
Return only a valid JSON object, no additional text:

{
  "query": "original query string",
  "annotations": [
    {
      "doc_id": "doc_id",
      "score": 0-3,
      "classification": "positive|hard negative|easy negative",
      "reasoning": "Brief, specific explanation of why the document does or does not satisfy the relevance definition. For hard negatives, pinpoint the plausible surface connection and the missing critical element."
    }
  ],
  "positive_doc_ids": ["list of doc_id with score 2 or 3"],
  "hard_negative_doc_ids": ["list of doc_id with score 1"],
  "easy_negative_doc_ids": ["list of doc_id with score 0"]
}
\end{Verbatim}
\Needspace{5\baselineskip}
\textbf{Complete refined relevance-annotation instruction (attempt 2):}\par
\begin{Verbatim}
You are a financial relevance annotator. For each query-document pair, judge whether the document substantially addresses the specific financial scenario, decision, or concept raised by the query and provides information that would materially help answer the user’s financial information need. Material help includes information that conclusively resolves the query—whether by confirming, refuting, or correcting the query’s underlying premise—as long as it directly engages with the core financial decision or scenario.

**Scoring Rubric (Absolute 0–3)**
- **Score 3 – Positive**: Directly and fully addresses the query’s precise financial scenario, decision, or concept. Contains the critical details, conditions, or actionable guidance the user actually needs, or provides a definitive resolution (including a clear, authoritative explanation that the proposed strategy is impossible, inadvisable, or based on a misunderstanding). No essential dimension is omitted.
- **Score 2 – Positive**: Adequately addresses the core need but is less complete, less specific, or less authoritative than a Score 3. It still provides material help—just missing some nuance, depth, or coverage of a secondary aspect. A document that supplies a largely correct answer but lacks full detail, or a refutation that omits some conditions, would fall here. Valid positive, but not the strongest answer.
- **Score 1 – Hard Negative**: Plausible at a glance (shares keywords, context, or the broad topic) but fails the relevance definition. It does **not** materially help answer the query because it either misses the specific financial scenario/decision, addresses a different angle, omits the decisive constraint, or provides information that is tangential, outdated, or inapplicable to the user’s situation. Only score 1 when a reasonable reader could initially think it might be relevant, but careful analysis shows it isn’t.
- **Score 0 – Easy Negative**: Clearly unrelated, off-topic, or outside the task’s relevance criteria. No plausible connection to the query.

**Hard Negative Reasoning (Score 1)**  
To identify hard negatives, ask: “What is the exact financial information need?” and “What makes this document look relevant, yet fail?”  
Common hard-negative patterns in this task:
- **Topic overlap, wrong decision context**: The document discusses the same general instrument or concept (e.g., “Roth IRA”), but the query asks about a specific strategy, tax treatment, or eligibility condition that the document never addresses (e.g., query: “tax implications of a backdoor Roth IRA” -> document explains the basics of a Roth IRA without mentioning backdoor conversions or their tax nuances).
- **Partial match missing the operative constraint**: The query includes a critical qualifier (e.g., “for self-employed,” “in retirement,” “for crypto”) that materially changes the answer. The document provides generic information without that qualifier, rendering the response inapplicable or misleading.
- **Wrong temporal or jurisdictional scope**: The document is about an outdated rule, a different country’s regulation, or a repealed provision when the query implies current, actionable guidance.
- **Looks helpful but solves the wrong problem**: The document explains how to perform an action, but the query asks whether the action is advisable in a specific context, or asks for comparison/alternatives the document never covers. **Caution**: If the document fully engages with the user’s scenario and provides a resolution (including a valid reason the action won’t work or is inadvisable), it solves the right problem and is relevant (Score 2 or 3). Do not mistake a corrective/dismissive answer for solving the wrong problem.

**Critical distinction—correction vs. mismatch**  
A document that conclusively addresses the query’s scenario by showing the proposed strategy is not feasible, not advisable, or based on a misunderstanding is **highly relevant**. It provides exactly the material help the user needs to make a correct financial decision. For example:  
- Query: “Can I use a lump-sum prepayment to move my next mortgage payment date forward?”  
- Document: “A lump-sum principal payment does not change your next payment due date; it only reduces the total interest and shortens the loan term.”  
This document directly and definitively answers the user’s uncertainty. It is a **positive** (Score 3 or 2), not a hard negative, because it engages with the exact scenario and provides a conclusive resolution.

If a document only tangentially mentions the topic or addresses a different aspect without engaging the core question, it may be a hard negative. Always check: does the document speak to the user’s stated financial decision or scenario, even if it contradicts the user’s initial assumption? If yes, it is relevant.

**Input Format**  
You will receive:  
- A query string (personal finance or investment question)  
- A list of candidate documents, each with a unique `doc_id`.

**Annotation Protocol**  
1. Process documents in the given order. Annotate every document exactly once; no duplication or omission.  
2. Evaluate each document independently. Do not compare documents; assign an absolute score based solely on the query-document relationship.  
3. Apply the rubric strictly. If in doubt between Score 1 and Score 2, re-read the query’s specific scenario/decision and check whether the document’s information materially advances an answer (including by definitively showing the proposed action is impossible or inadvisable).

**Output Format**  
Return only a valid JSON object, no additional text:

{
  "query": "original query string",
  "annotations": [
    {
      "doc_id": "doc_id",
      "score": 0-3,
      "classification": "positive|hard negative|easy negative",
      "reasoning": "Brief, specific explanation of why the document does or does not satisfy the relevance definition. For hard negatives, pinpoint the plausible surface connection and the missing critical element. For positives that refute the query’s premise, note that it directly addresses and resolves the core scenario."
    }
  ],
  "positive_doc_ids": ["list of doc_id with score 2 or 3"],
  "hard_negative_doc_ids": ["list of doc_id with score 1"],
  "easy_negative_doc_ids": ["list of doc_id with score 0"]
}
\end{Verbatim}
\clearpage
\phantomsection\addcontentsline{toc}{section}{Supplementary Note 2: Experimental configurations}
\section*{Supplementary Note 2: Experimental configurations}
\textbf{Test-mode models.} A commonly used test-mode configuration uses GPT-4.1-mini to evaluate identifier--attribute outputs and refine their generation guidance. Gemini 3 Flash Preview generates, evaluates and refines task definitions and query-generation instructions, and constructs and refines relevance-annotation instructions using feedback from programmatic few-shot checks. Qwen3-30B-A3B-Instruct-2507 generates identifiers, attributes, queries and relevance labels in both test and production modes.

Supplementary Table~\ref{tab:si-training-settings} summarizes the shared training settings for the self-improvement comparison, component ablations and MS MARCO query-source comparison.
Supplementary Table~\ref{tab:si-strong-scaling-settings} gives the settings for adaptation of strong embedding models and the MS MARCO data-scaling experiment.

\begingroup\small
\begin{longtable}{P{0.31\textwidth}P{0.64\textwidth}}
\caption{Shared training settings for the self-improvement comparison, component ablations and MS MARCO query-source comparison.}\label{tab:si-training-settings}\\
\toprule Parameter & Setting \\ \midrule\endfirsthead
\multicolumn{2}{l}{Supplementary Table \thetable\ (continued)}\\
\toprule Parameter & Setting \\ \midrule\endhead
\bottomrule\endfoot
Initialization & Qwen3-0.6B-auto-eos \\
Pooling & Last token \\
Embedding normalization & L2 normalization \\
LoRA configuration & Rank 32, alpha 64, dropout 0.1 \\
Learning rate & $10^{-4}$ \\
Global query batch size & 96 \\
Training epochs & 1 \\
Warmup ratio & 0.1 \\
Random seed & 42 \\
\end{longtable}\endgroup

\begingroup\small
\begin{longtable}{P{0.25\textwidth}P{0.22\textwidth}P{0.22\textwidth}P{0.23\textwidth}}
\caption{Training settings for adaptation of strong embedding models and the MS MARCO data-scaling experiment. The strong-model settings apply to LeCaRDv2, EvidenceBench, SciRepEval Search, TheoremQA questions, SpartQA and WinoGrande.}\label{tab:si-strong-scaling-settings}\\
\toprule Parameter & BGE-M3 & Qwen3-Embedding-8B & Qwen3-8B-auto-eos (scaling) \\ \midrule\endfirsthead
\multicolumn{4}{l}{Supplementary Table \thetable\ (continued)}\\
\toprule Parameter & BGE-M3 & Qwen3-Embedding-8B & Qwen3-8B-auto-eos (scaling) \\ \midrule\endhead
\bottomrule\endfoot
Learning rate & $10^{-5}$ & $10^{-4}$ & $10^{-4}$ \\
Training epochs & 1; 5 for SpartQA & 1 & 1 \\
Query batch size per GPU & 2; 16 for SpartQA & 16 & 16 \\
Pooling & CLS token & Last token & Last token \\
Embedding normalization & L2 & L2 & L2 \\
Maximum query / passage length (tokens) & 512 / 512 & 512 / 512 & 512 / 512 \\
Warmup ratio & 0.1 & 0.1 & 0.1 \\
Random seed & 42 & 42 & 42 \\
\end{longtable}\endgroup
For BGE-M3 on SpartQA, we increased the number of training epochs and the batch size because preliminary runs indicated insufficient training under the default settings. Both Qwen3 models in Supplementary Table~\ref{tab:si-strong-scaling-settings} use LoRA with rank 32, alpha 64 and dropout 0.1. For tasks with existing labelled training data, the original-data and original-plus-SYNTRA conditions use the same training settings. The scaling experiment uses the same training settings and evaluation protocol across all six data scales (10,000, 20,000, 40,000, 80,000, 160,000 and 320,000).

\textbf{Human-evaluation sampling.} For the seven-annotator relevance-label evaluation, we randomly sampled 50 query--document pairs without replacement from the LLM-annotated candidate pairs for each of MS MARCO, Text2SQL and TheoremQA (theorems), yielding 150 pairs in total.